\documentclass[fleqn,usenatbib]{mnras}

\usepackage{newtxtext,newtxmath}
\usepackage{longtable}
\usepackage[table,xcdraw]{xcolor}
\usepackage[T1]{fontenc}
\DeclareRobustCommand{\VAN}[3]{#2}
\let\VANthebibliography\thebibliography
\def\thebibliography{\DeclareRobustCommand{\VAN}[3]{##3}\VANthebibliography}
\usepackage{graphicx}	% Including figure files
\usepackage{amsmath}	% Advanced maths commands
\usepackage{gensymb} % For \degree symbol
\usepackage{cleveref}
\usepackage{caption}

\title[ATA FXT Follow-up]{Radio Follow-Up of Einstein Probe Fast X-Ray Transients}

\author{
Carmen Choza$^{1,4}$,
Joe S. Bright$^{1,3,4}$, Francesco Carotenuto$^{2}$, Alex Pollak$^{3}$, Rob Fender$^{1}$, and Andrew Siemion$^{1,3,4}$
\\
$^{1}$Astrophysics, Department of Physics, The University of
Oxford, Keble Road, Oxford, OX1 3RH, UK.\\
$^{2}$INAF Osservatorio Astronomico di Roma, Via Frascati 33, I-00078, Monte Porzio Catone (RM), Italy\\
$^{3}$SETI Institute, 339 Bernardo Ave., Suite 200 Mountain View, CA 94043, USA\\
$^{4}$Breakthrough Listen, University of Oxford, Department of Physics, Denys Wilkinson Building, Keble Road, Oxford, OX1 3RH, UK\\
}

\date{Accepted XXX. Received YYY; in original form ZZZ}

\pubyear{\the\year{}}

\begin{document}
\label{firstpage}
\pagerange{\pageref{firstpage}--\pageref{lastpage}}
\maketitle

% Abstract of the paper
\begin{abstract}
Fast X-ray transients (FXTs) are brief, luminous bursts of soft X-ray emission whose physical origins remain uncertain. The Einstein Probe (EP) mission has recently enabled prompt discovery of these events, providing opportunities for rapid multi-wavelength follow-up. We present a coordinated radio observing campaign targeting 20 FXTs detected by the EP in 2024. The core consists of 59 epochs with the Allen Telescope Array (ATA), sampling post-burst timescales from $\sim$1 to $\sim$65 days and reaching typical $3\sigma$ sensitivities of 0.6--1.5\,mJy across 1--8\,GHz. We additionally incorporate higher-sensitivity observations from MeerKAT, the VLA, ATCA, eMERLIN, and AMI-LA. Two FXTs---EP240315a and EP241021a---have radio counterparts. For EP241021a, AMI-LA monitoring at 15.5\,GHz reveals a light curve peaking at $\sim$30 days with $F_{\nu}\approx1.0$\,mJy; equipartition analysis implies a Newtonian equipartition energy of $\sim3\times10^{49}$\,erg and a mildly relativistic on-axis solution with $\Gamma_{\rm on}\approx1.3$. Multi-frequency modelling of EP240315a, combining new VLA measurements with published MeerKAT, ATCA, and eMERLIN data, indicates highly relativistic early emission ($\Gamma\gtrsim3$) that decelerates with time, consistent with jetted outflow. The detections and ATA non-detections span luminosities overlapping the brightest GRB and relativistic tidal disruption event (TDE) afterglows, suggesting FXTs comprise a heterogeneous population including both relativistic and non-relativistic explosions. Under a GRB-like redshift prior, our $3\sigma$ ATA limits correspond to observer-frame specific-luminosity thresholds of $\sim10^{32}$--$10^{33}\,\mathrm{erg\ s^{-1}\ Hz^{-1}}$; for nearby events ($z\lesssim0.2$), $L_{\nu}\lesssim10^{30\text{--}31}\,\mathrm{erg\ s^{-1}\ Hz^{-1}}$. Planned ATA upgrades and next-generation arrays (SKA, DSA, ngVLA) will enable sensitive, population-level radio studies of the FXT radio sky.
\end{abstract}


\section{Introduction}

% This works for a beginning; read recent literature to see if you want to add any information on what FXTs are 

Astronomical observations have revealed enigmatic phenomena known as Fast X-ray Transients (FXTs), characterized by brief bursts of soft X-ray radiation that persist for durations ranging from tens to thousands of seconds \citep{quirola-vasquez_extragalactic_2022, quirolavasquez2023}. Until recently, these events were primarily identified through retrospective analysis of archival data from major X-ray observatories, including Chandra, XMM-Newton, \textit{Swift}, and eROSITA \citep{Jonker2013,Glennie2015}. This delayed discovery process significantly limited opportunities for immediate multi-wavelength observations, with a few exceptions noted in the literature \citep{soderbergExtremelyLuminousXray2008a, ibrahimzade_constraints_2025}.
 % Find other significant literature
The origin of FXTs remains a subject of active investigation, with several theoretical models proposed to explain their X-ray signatures. These include the disruption of white dwarfs by tidal forces \citep{Jonker2013,Glennie2015}, intense stellar flare events \citep{Glennie2015}, the initial breakout of supernova shocks \citep{soderbergExtremelyLuminousXray2008a,Novara_2020,2020ApJ...896...39A}, long-duration gamma-ray bursts \citep{Jonker2013,bauer2017}, and emissions from newly formed, rapidly rotating neutron stars with strong magnetic fields \citep{Xue2019}. The extraordinary range in observed luminosities, spanning from $10^{30}\,\rm{ergs}\,\rm{s}^{-1}$ for potential stellar flares \citep{Glennie2015} to $10^{48}\,\rm{ergs}\,\rm{s}^{-1}$ for distant cosmic events \citep{Srivastav2025}, implies multiple underlying mechanisms are almost certainly responsible. However, definitive conclusions have been hampered by the challenges in identifying host galaxies for extragalactic FXTs \citep{Eappachen2022,quirolavasquez2023}. 

The launch of the Einstein Probe X-ray Telescope (EP; \citealt{yuanEinsteinProbeMission2022}) in early 2024 has triggered a rapid expansion of the discovery and study of FXTs. The Wide-field X-ray Telescope on the EP (EP-WXT) is sensitive in the soft X-ray regime between 0.5–4\,keV, and its wide (3600 deg$^2$) field of view provides efficient sky-monitoring; the secondary instrument on the probe, the Fast X-ray Telescope (EP-FXT), provides rapid follow-up and localisation on the order of 3\arcsec. Detections of candidate extragalactic FXTs are released with low latency via NASA's General Coordinates Network (GCN; single alerts also referred to as GCNs), for an average of about 2 sources per week in 2024. The sample continues to increase rapidly, providing the opportunity for timely multi-wavelength follow-up and monitoring campaigns. Prompt follow-up has resulted in the detection of several optical and radio counterparts, and early EP detections have already demonstrated diverse behaviour: some show strong multi-wavelength counterparts and clear associations with known astrophysical phenomena, while many others lack any prompt gamma-ray association, implying either off-axis, low-luminosity relativistic outflows or inherently different explosion channels \citep[see e.g.][]{pugliese2024,shuEP241021a,busmann_curious_2025}. As of the end of 2024, only eleven had demonstrated prompt optical counterparts sufficiently luminous to enable calculation of spectroscopic redshifts; as of late August 2025, $\sim$\,25\% of Einstein Probe FXTs have established spectroscopic redshifts, with a smaller sample of additional redshifts determined through association with a host galaxy \citep{oconnorRedshiftDistributionEinstein2025}.

Sources with spectroscopic redshifts and luminous optical counterparts have been of particular interest to the community. The majority have occurred at redshifts below $z \approx 2$, with significant exceptions \citep{oconnorRedshiftDistributionEinstein2025}. One notable exception is EP240315a, a highly luminous source at z = 4.8 with the first detected radio afterglow \citep{gillanders2024}. Unlike the majority of EP-FXT sources \citep{yadav_radio_2025}, EP240315a was associated with a simultaneous GRB \citep{svinkinKonusWindDetectionGRB2024}. EP240315a was the target of substantial observation and analysis, determining it to be the result of a highly relativistic, jetted outflow associated to a long GRB event \citep{gillanders2024,levanFastXrayTransient2025}. This work combines results of follow-up observations of the radio counterpart of EP240315a with detections in the literature \citep{carotenuto2024, gillanders2024,liu2024}. A contrasting example is EP241021a, which exhibits no detectable gamma-ray counterpart \citep{svinkin2024,Burns2024}, and was localized to a redshift of z = 0.75 \citep{pugliese2024}, ruling out the probability of a distant GRB peak redshifted into the X-ray. Despite the lack of a gamma-ray counterpart, EP241021a exhibited a bright optical counterpart \citep{busmann_curious_2025,shuEP241021a} and a long-lived radio afterglow \citep{yadav_radio_2025,gianfagnaSoftXrayTransient2025a}. This work presents radio monitoring of EP241021a at 15.5\,GHz using AMI-LA, spanning from $\sim$10 to $\sim$150 days post-FXT trigger in the observer frame.

Although the diversity of proposed FXT origins spans a wide range of physical scenarios, direct constraints on their dynamics and energy release remain scarce. The rapid, well-localized detections enabled by the Einstein Probe now make it possible to investigate these transients across the electromagnetic spectrum from their earliest phases. Among the available follow-up wavelengths, radio observations are uniquely sensitive to the kinetic energy and geometry of the outflow, tracing synchrotron emission from the interaction of ejecta with the surrounding medium. By monitoring this emission over weeks to months, one can measure expansion velocities, probe energy coupling to the environment, and distinguish between jetted, mildly relativistic, and non-relativistic explosions. In this way, radio studies provide a direct means of connecting FXTs to known classes such as GRBs and TDEs—or revealing a population with distinct physical origins.

In this paper, we present the results of a radio observing campaign of 20 Einstein Probe FXTs conducted with the Allen Telescope Array. We also present the results of additional monitoring of EP240315a with MeerKAT and the Karl G. Jansky Very Large Array (VLA) and of EP241021a with the Arcminute Microkelvin Imager Large Array (AMI-LA) and analysis of the radio light curves in combination with results from the literature \citep{carotenuto2024,gillanders2024,liu2024,ricciLongtermRadioMonitoring2025}. In \Cref{sec:observations} we describe the EP sample, the GCN-based trigger and scheduling strategy, and the telescopes, observing setups and reduction pipelines used in our campaign. \Cref{sec:observations} also details our imaging, flux-extraction and upper-limit procedures and presents the full radio dataset (flux densities and $3\sigma$ limits). In \Cref{sec:results} we model the two well-sampled radio counterparts (EP240315a and EP241021a): multi-frequency light-curve fitting, broken-power-law modelling, and equipartition-based estimates of radius, energy and apparent expansion speed are presented, together with a grid search of relativistic jet parameter space, and additionally gives a survey-level overview of detections and non-detections and compares our ATA limits to representative radio light curves of GRBs, relativistic TDEs and other classes. \Cref{sec:discuss} discusses implications for the nature of FXTs — energetics, Lorentz factors, viewing-angle interpretations, and population constraints under a GRB-like redshift prior — and \Cref{sec:facilities} outlines survey sensitivity, prospects for future facilities (upgraded ATA, SKA, DSA, ngVLA), and observational priorities to improve host identification and multi-wavelength characterization. We adopt a flat cosmology (H$0=70\,$km\,s$^{-1}$\,Mpc$^{-1}$ $T_{\rm CMB}=2.725,$K, $\Omega_\Lambda=0.3$) throughout; distances are computed using the \texttt{astropy.cosmology} package.

\section{Observations}\label{sec:observations}

Here we present the FXT sample and the reduction processes for data collected at respective facilities. The times of discovery and positions of the sources for our survey were collected from the GCN alert stream. The sample is presented in Table 1, with flux density measurements and upper limits presented in Appendix A.

\begin{table*}

\caption{Fast X-ray transient properties and times of initial discovery for the 20 sources observed in this work. A subset of sources have been confirmed to have optical counterparts; we report spectroscopic redshifts for those sources. References: (a) \citealt{levanFastXrayTransient2025}, (b) \citealt{Srivastav2025}, (c) \citealt{quirolavasquez2024}, (d) \citealt{quirolavasquez2024b}, (e) \citealt{pugliese2024}.}
\label{tab:ata_sample}
\begin{tabular}{llccc}
\textbf{Einstein Probe ID}                & \multicolumn{1}{c}{\textbf{Date of discovery}}         & \multicolumn{1}{c}{\textbf{RA\_EP (deg)}} & \multicolumn{1}{c}{\textbf{Dec\_EP (deg)}} & \textbf{Redshift (z)} \\ 

\hline
EP240315a  &  2024-03-15 20:10:44 & 141.644   & -9.547     & 4.859 (a) \\
EP240413a                        &  2024-04-13 14:39:37 &  228.794   &  -18.8      & - \\
EP240414a                        &  2024-04-14 9:50:12  &  191.498   &  -9.695     & 0.4016 (b) \\
EP240416a                        &  2024-04-16 2:42:13  &  203.15    &  -13.612    & - \\
EP240417a                        & 2024-04-17 15:12:33  & 177.442    & -15.438     & - \\
EP240420a                        & 2024-04-20 12:04:28 &  228.713   &  14.796     & - \\
EP240506a                        & 2024-05-06 5:01:39                         & 213.978                          & -16.715                           & - \\
EP240618a                        & 2024-06-18 5:43:43                         & 281.627                          & 23.82                             & - \\
EP240625a                        & 2024-06-25 1:48:23                         & 310.76                           & -15.966                           & - \\
EP240626a                        & 2024-06-26 6:28:28                         & 263.023                          & -13.051                           & - \\
EP240703a                        & 2024-07-03 0:38:40                         & 273.803                          & -9.681                            & - \\
EP240708a                        & 2024-07-08 23:28:23                        & 345.963                          & -22.84                            & - \\
EP240801a                        & 2024-08-01 9:06:03                         & 345.14                           & 32.61                             & 1.673 (c) \\
EP240806a                        & 2024-08-06 4:47:54                         & 11.491                           & 5.091                             & 2.818 (d) \\
EP240816b                        & 2024-08-16 1:44:27                         & 16.0161                          & 15.4151                           & - \\
EP240908a                        & 2024-09-08 18:49:05                        & 14.0031                          & 8.0735                            & - \\
EP240913a                        & 2024-09-13 11:39:33                        & 16.681                           & 16.75                             & - \\
EP240918a                        & 2024-09-18 11:21:52                        & 289.3933                         & 46.1278                           & - \\
EP240919a                        & 2024-09-19 14:53:43                        & 334.2797                         & -9.7362                           &  - \\
EP241021a                        & 2024-10-21 5:07:56                         & 28.852                           & 5.957         & 0.748 (e)     \\
\hline
\end{tabular}
\end{table*}

\subsection{ATA}\label{section:ATA}

The Allen Telescope Array (ATA) is a 43-element radio interferometer hosted at the Hat Creek Radio Observatory (Pollak et al., in prep). Each active antenna is equipped with log-periodic dual polarization feeds that are tunable in the range of 1 to 12\,GHz. The correlator backend supports digitization of four simultaneous tunings\footnote{Two before February 2025; this survey consists of a maximum of two bands per observation.}, where each tuning can be assigned a discrete central frequency anywhere within the available frequency range of the feeds. Each tuning records a bandwidth of 672\,MHz with flexible frequency and time resolution. Before June $4^{\text{th}}$ 2024, the array consisted of 21 active antennae; after June $4^{\text{th}}$ the number of active antennae increased to 28. Accordingly, our observations after that date place stricter average constraints because of the higher sensitivity of the instrument. Spectroscopic redshifts have been determined for five of the sources in our sample and are reported in \Cref{tab:ata_sample}. Of those, three have detected radio counterparts: EP240315a \citep{gillanders2024}, EP240414a \citep{brightRadioCounterpartFast2025}, and EP241021a \citep{huEP241021aEPDetection2024}.

A total of 20 FXT sources were observed as part of the ATA observing campaign. From March 2024 to October 2024, GCNs were monitored daily, and telescope time was scheduled as rapidly as possible to follow X-ray transient alerts. As a result, half of the targets observed in this campaign were observed within one week of initial detection with the EP, and all but one were observed at least once within a month (with the exception of some extended observations, which represented targets revisited for longer-term monitoring). Initially, observations were planned to collect long (many hour) stares at a range of RF bands spanning 1 - 8\,GHz. Observations prioritized frequency coverage, recording at four bands (centred at 1500\,MHz, 3000\,MHz, 5000\,MHz, and 8500\,MHz) over the course of an observation, beginning with the lower pair of spectral bands and transferring to the second pair later. During the second half of the campaign, the observing strategy was adjusted to place the two simultaneous spectral windows adjacently to maximize sensitivity by enabling combined imaging across the full recording bandwidth (with central frequencies of 2660\,MHz and 3340\,MHz at lower frequencies, and 6660\,MHz and 7340\,MHz at higher frequencies). Observations at each set of frequencies were scheduled as close together in time as possible. Where possible, observations were scheduled to ensure at least 2 hours on source. Observations vary in duration from 0.5 to 5 hours, for an average of $\sim$2.6 hours per observation per spectral window.

Observations were reduced with a \texttt{CASA}-based calibration pipeline\footnote{https://github.com/joesbright/ATARI.git} and imaged using \texttt{wsclean} \citep{McMullin2007,Offringa2014,casateamCASACommonAstronomy2022}. Flux scales, gain, and bandpass solutions were obtained using one of 3C286, 3C48, or 3C147, with complex gain calibrators chosen to be nearby (within 10 degrees) unresolved sources in the Very Large Array (VLA) calibrator catalogue with brightnesses above 1\,Jy at the relevant spectral bands. Flux calibrators were observed at the beginning of observing blocks, and thereafter the FXT target positions were observed interleaved with 5-minute scans of the complex gain calibrator. 

In total, we collected 59 observations, for an average of 4.5 hours per source across the radio spectrum. We find no detections; for each observation, 3-$\sigma$ upper limits were calculated from the root mean square (RMS) noise in the residual maps. Limits per spectral window, as well as the limits of combined imaging across the full simultaneous bandwidth, are presented in \Cref{app:limits}.

\subsection{MeerKAT}

We first observed the field of EP240315a between UT 16:35:59 and UT 17:24:12 on 2024-03-18 using the S4 tuning of the S-band receiver (with a central frequency of $3.06\,\rm{GHz}$) under project code SCI-20230907-JB-01 (P.I. Bright). The total bandwidth for the observation was 875\,MHz and PKS~J1939$-$6342 and 3C237 (J1008$+$0730) were used as flux/bandpass and complex gain calibrators, respectively. Data calibration and imaging were performed using the \texttt{OxKAT} pipeline \citep{oxkat}. This observation led to the discovery of the radio counterpart to EP240315a \citep{carotenuto2024,gillanders2024} at $34\pm5\,\muup\rm{Jy}$. All further observations of EP240315a with MeerKAT were observed and reduced in the same manner, and are reported for the first time in this work. When reporting errors, we add a 5\% flux scale uncertainty in quadrature to the statistical uncertainty derived when fitting the flux density \citep{knowles2022}. Our MeerKAT observations of EP240315a are summarized in \Cref{tab:radio_data}.

% Section on ep240315a? Need to save the discussion of this object in context until later, 
In addition to our observations with MeerKAT and the VLA we include ATCA observations presented in \citet{liu2024} and \citet{ricciLongtermRadioMonitoring2025} and eMERLIN observations from \cite{gillanders2024} to supplement our dataset.

\begin{table}
	\centering
	\caption{Radio observations of EP240315a included in this work and reproduced from the literature. Upper limits are given at the 3-sigma level. The reference time with which $\Delta T$ is measured with respect to is UTC 20:10:14 on 15 March 2024 (MJD 60384.840) \citet{liu2024}. Note that the data corresponding to the two eMERLIN non-detections were combined to detect the source at $\Delta T=8.550$. References: (1) \citet{carotenuto2024}, (2) \citet{gillanders2024}, (3) \citet{liu2024}, \citet{ricciLongtermRadioMonitoring2025}.}
	\label{tab:radio_data}
	\begin{tabular}{ccccc}
		\hline
		  $\Delta T$ & Flux Density & Facility & Frequency & Reference\\
            Days & $\muup\rm{Jy}$ & & GHz &\\
		\hline
            2.87 & $34\pm5$ & MeerKAT & 3 & (1,2) \\
            
            3.951 & $74\pm21$ & ATCA & 5.5 & (3) \\
            3.951 & $82\pm20$ & ATCA & 9.0 & (3) \\
            
            5.066 & $<75$ & eMERLIN & 5 & (2,3) \\
            
            8.550 & $67\pm13$ & eMERLIN & 5 & (2,3) \\
            
            10.524 & $97\pm14$ & ATCA & 5.5 & (3) \\
            10.524 & $158\pm12$ & ATCA & 9.0 & (3) \\
            
            11.256 & $150\pm7$ & VLA & 9.0 & This work \\
            11.256 & $140\pm10$ & VLA & 11.0 & This work \\
            
            11.279 & $160\pm13$ & VLA & 12.4 & This work \\
            11.279 & $180\pm10$ & VLA & 13.2 & This work \\
            11.279 & $180\pm10$ & VLA & 13.9 & This work \\
            11.279 & $170\pm10$ & VLA & 14.7 & This work \\
            11.279 & $200\pm10$ & VLA & 16.2 & This work \\
            11.279 & $180\pm10$ & VLA & 17 & This work \\
            
            12.034 & $<105$ & eMERLIN & 5 & (2,3) \\
            
            19.576 & $140\pm27$ & ATCA & 5.5 & (3) \\
            19.576 & $328\pm31$ & ATCA & 9.0 & (3) \\
            
            25.85 & $66\pm6$ & MeerKAT & 3.0 & This work \\

            26.288 & $93\pm8$ & VLA & 9.0 & This work \\
            26.288 & $100\pm10$ & VLA & 11.0 & This work \\

            26.310 & $90\pm20$ & VLA & 12.8 & This work \\
            26.310 & $100\pm10$ & VLA & 14.3 & This work \\
            26.310 & $100\pm10$ & VLA & 15.9 & This work \\
            26.310 & $100\pm10$ & VLA & 17.4 & This work \\
            
            64.95 & $64\pm5$ & MeerKAT & 3.0 & This work \\
		\hline
	\end{tabular}
\end{table}

\subsection{VLA}

We were awarded directors discretionary time to observe EP240315a under the project code 24A-455 (P.I. Bright). Our first observation began on 2024-03-27 at UT 02:05:14.300 using the X-band receiver, immediately followed by an observation with the Ku-band receiver, with the WIDAR correlator set up for maximum bandwidth in both instances. We observed 3C147 to calibrate the instrument bandpass and set the absolute flux scale for both bands. The compact source J0943$-$0819 was used to calibrate the time-dependent complex gains of the antennas at both observing bands.

Data were reduced using the VLA data reduction pipeline with flagging applied according to the recommendations of the Science Ready Data Products initiative \citep{sdrp2019}. We then averaged the data to $16\,\rm{Hz}$ channel widths and $9\,\rm{s}$ correlator integration times to reduce the computational cost of imaging. Such averaging will produce minimal image distortions due to bandwidth and temporal smearing (see EVLA memo 199 \footnote{https://library.nrao.edu/public/memos/evla/EVLAM\_199.pdf}). Data were imaged using the \texttt{tclean} task in CASA using multi-frequency synthesis with two Taylor terms and we split the observing bandwidth into smaller chunks when imaging in order to extract more spectral information. When reporting errors we add a 5\% flux scale uncertainty in quadrature to the statistical uncertainty derived when fitting the flux density as advised by the VLA Observational Status Summary 2024B\footnote{https://science.nrao.edu/facilities/vla/docs/manuals/oss/performance/fdscale}.

\subsection{AMI-LA}
The Arcminute Microkelvin Imager Large Array (AMI-LA) is an eight element interferometer operating at a central frequency of $15.5\,\rm{GHz}$ across a $5\,\rm{GHz}$ bandwidth \citep{zwart2008,2018MNRAS.475.5677H} which performs regular transient monitoring. Observations were performed of three FXTs, EP240618A, EP240801A, and EP241021a, with the observations summarised in \Cref{tab:AMI-LA}. Data were reduced using the custom \textsc{reduce\_dc} pipeline which works with a bespoke AMI-LA data product where data are channelised into 8 coarse channels of $625\,\rm{MHz}$ width. The flux scale was set with the regular flux density calibrator 3C286 and phase reference calibration was performed from a nearby bright and compact source, J1850$+$2825, J2236$+$2828, and J0149$+$0555, respectively. 

% Detected lightcurve points
\begin{table}
	\centering
	\caption{Radio observations of EP241021a, EP240618a, and EP240801a obtained with AMI-LA and included in this study. We list 3 sigma limits for those epochs where the source was not detected. Flux errors for detections include statistical and absolute calibration uncertainties.}
    \label{tab:AMI-LA}
    \begin{tabular}{ccccc}
    \hline
    \textbf{Date} & \textbf{Flux}  & \textbf{Flux Error}  & $\Delta$\textbf{T} \\
    \textbf{(yy-mm-dd)} & ($\mu$\textbf{Jy}) & ($\mu$\textbf{Jy}) & \textbf{(days)} \\
    \hline
    \textbf{EP240618A}      &                    &                   & \\
    2024-06-18     & \textless\,120   & -     & 0.73          \\
    2024-06-22     & \textless\,114   & -     & 3.83          \\
    \hline
    \textbf{EP240801A}      &                    &                   & \\
    2024-08-09$^2$     & \textless\,133  & -      & 7.78          \\
    2024-08-19$^2$   & \textless\,148  & -      & 17.64        \\
    \hline
    \textbf{EP241021A}      &                    &                   & \\
    2024-10-30      & 488          & 29                 & 9.78          \\
    2024-10-31      & 532          & 40                 & 10.74         \\
    2024-11-02      & 745          & 154                & 12.75         \\
    2024-11-05      & 865          & 67                 & 15.74         \\
    2024-11-06      & 1010         & 77                 & 16.78         \\
    2024-11-07      & 876          & 54                 & 17.73         \\
    2024-11-09      & 844          & 44                 & 18.81         \\
    2024-11-10      & 850          & 44                 & 20.72         \\
    2024-11-11      & 837          & 42                 & 21.72         \\
    2024-11-13      & 1070         & 60                 & 23.73         \\
    2024-11-15      & 1100         & 59                 & 25.68         \\
    2024-11-17      & 1150         & 67                 & 27.77         \\
    2024-11-27      & 1020         & 54                 & 37.74         \\
    2024-12-06      & 1030         & 60                 & 46.62         \\
    2024-12-09      & 809          & 49                 & 49.65         \\
    2024-12-13      & 839          & 46                 & 53.70          \\
    2024-12-16      & 885          & 49                 & 56.69         \\
    2024-12-23      & 630          & 40                 & 63.62         \\
    2024-12-25      & 705          & 36                 & 65.66         \\
    2024-12-27      & 580          & 37                 & 67.59         \\
    2024-12-28      & 638          & 58                 & 68.59         \\
    2024-12-29      & 673          & 48                 & 69.58         \\
    2025-01-02      & 529          & 46                 & 73.58         \\
    2025-01-28    & \textless\,349  & -          & 99.53         \\
    2025-02-06    & \textless\,251  & -          & 108.5         \\
    2025-02-15    & \textless\,430  & -          & 117.5         \\
    2025-02-18    & \textless\,242  & -          & 120.5         \\
    2025-03-05    & \textless\,157  & -          & 135.4         \\
    2025-03-06      & 568          & 64                 & 136.5        \\
    2025-03-10    & \textless\,338  & -          & 140.4         \\
    2025-03-17    & \textless\,335  & -          & 147.3   \\    
    
    \hline \\
    \end{tabular}
\end{table}

\section{Results}\label{sec:results}

\subsection{Upper Limits from the ATA}

\begin{figure*}
    \includegraphics[width=\textwidth]{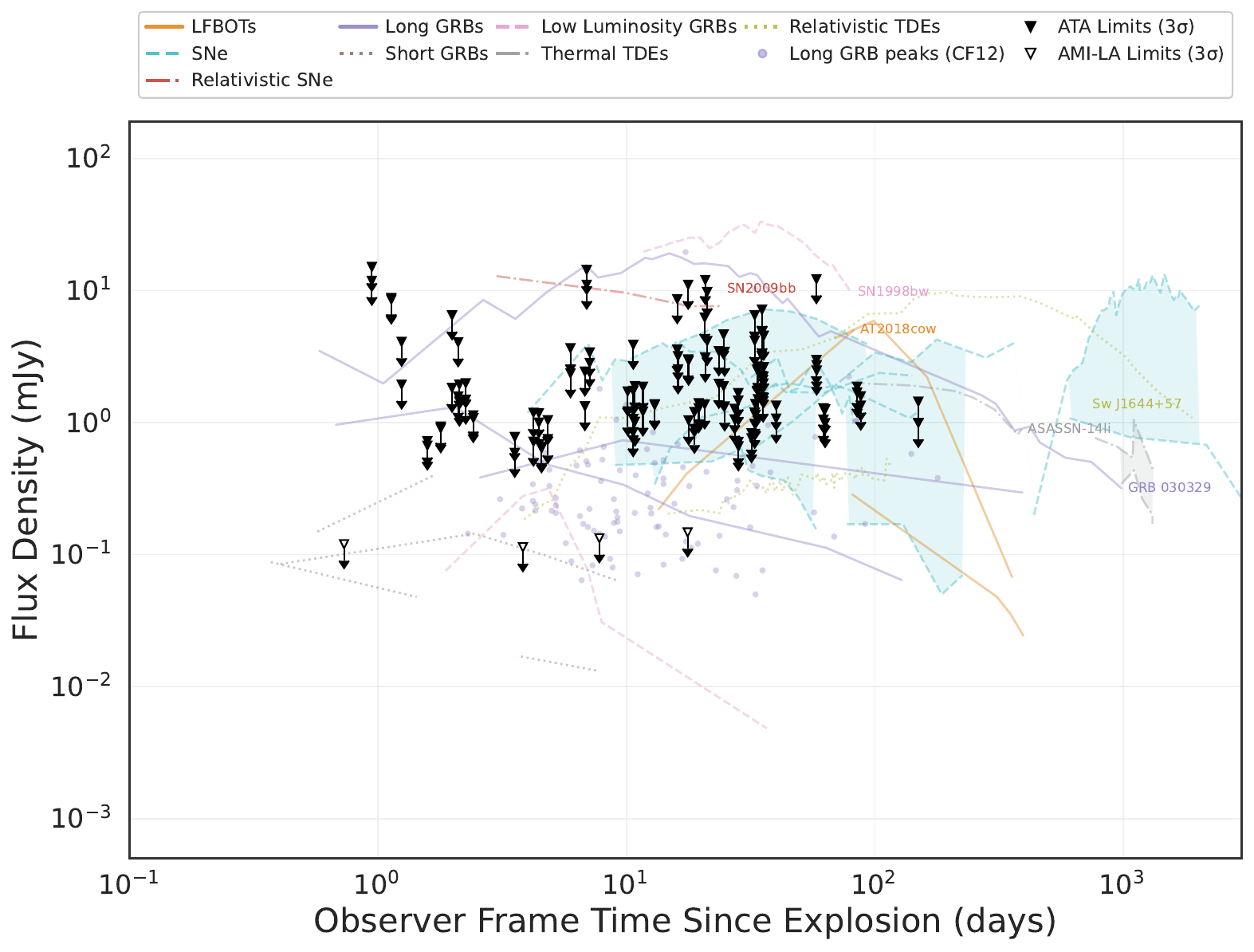}
    \caption{The population of $3\sigma$ upper limits derived from our ATA observations (filled black triangles, with downward arrows for visibility), plotted against radio light curves of extragalactic transient classes in flux density: supernovae, relativistic supernovae, LFBOTs, thermal TDEs, relativistic TDEs, and short, long, and low-luminosity GRBs. AMI-LA upper limits for two sources are shown as hollow triangles (see \Cref{tab:AMI-LA}). Shaded bands span the minimum--maximum flux of all plotted members of the supernova sample of \citet{bietenholzRadioLuminosityrisetimeFunction2021} and the tidal disruption event sample of \citet{cendesUbiquitousLateRadio2024}; small purple circles mark the peak flux densities and peak times of the long-GRB radio afterglows of \citet{chandra_radio-selected_2012}. All comparison data are taken at observing frequencies of 1--10\,GHz, comparable to our ATA band. Comparisons are shown in flux space---i.e., without luminosity scaling---because most EP sources lack redshift measurements; times are also in the observer frame. These limits indicate that radio counterparts as bright as the most luminous events found to date may be uncommon in the broader population. This figure, and subsequent figures including comparison light curves, are based on those presented in \citet{ho2020} and \citet{brightRadioCounterpartFast2025}, with additional light curves drawn from \citet{chandra_radio-selected_2012}, \citet{bietenholzRadioLuminosityrisetimeFunction2021}, and \citet{cendesUbiquitousLateRadio2024}.}
    \label{fig:ata_flux_limits}
\end{figure*}

Considering first just the flux density upper limits of the EP FXT population as constrained by the ATA we see the majority of our limits clustering at the $\sim1\,\rm{mJy}$ level (\Cref{fig:ata_flux_limits}). This is broadly consistent with the detection and upper-limit distributions reported by \citet{chandra_radio-selected_2012} (their Table~2 and Fig.~4), who found that most GRB afterglows with a detected radio counterpart had a flux density between the $\sim\, \muup$Jy level and $\sim$1~mJy at 8.5\,GHz; although the ATA observations are at different frequencies, our upper limits occupy a comparable flux-density regime. The limits sample the early to intermediate post-explosion phases in the observer frame, spanning $\simeq1$–$65$~days; our earliest constraint was obtained at $t_{\rm obs}\approx0.94$~days after the explosion, and our deepest single-epoch limit is $F_{\nu}\approx0.6$~mJy (3$\sigma$). All limits are displayed in \Cref{fig:ata_flux_limits} in comparison to a sample of radio counterparts of other extragalactic transients. Individual limits at each frequency band and epoch are reported in \Cref{app:limits}. The AMI-LA observations of EP240618a and EP240801a did not yield detections. Upper limits are reported at the 3$\sigma$ level in \Cref{tab:AMI-LA} and are at the $\sim100\,\muup\rm{Jy}$ level.

% Detections, stating info on the sources we do see
\subsection{EP241021a}

\begin{figure}
	\includegraphics[width=\columnwidth]{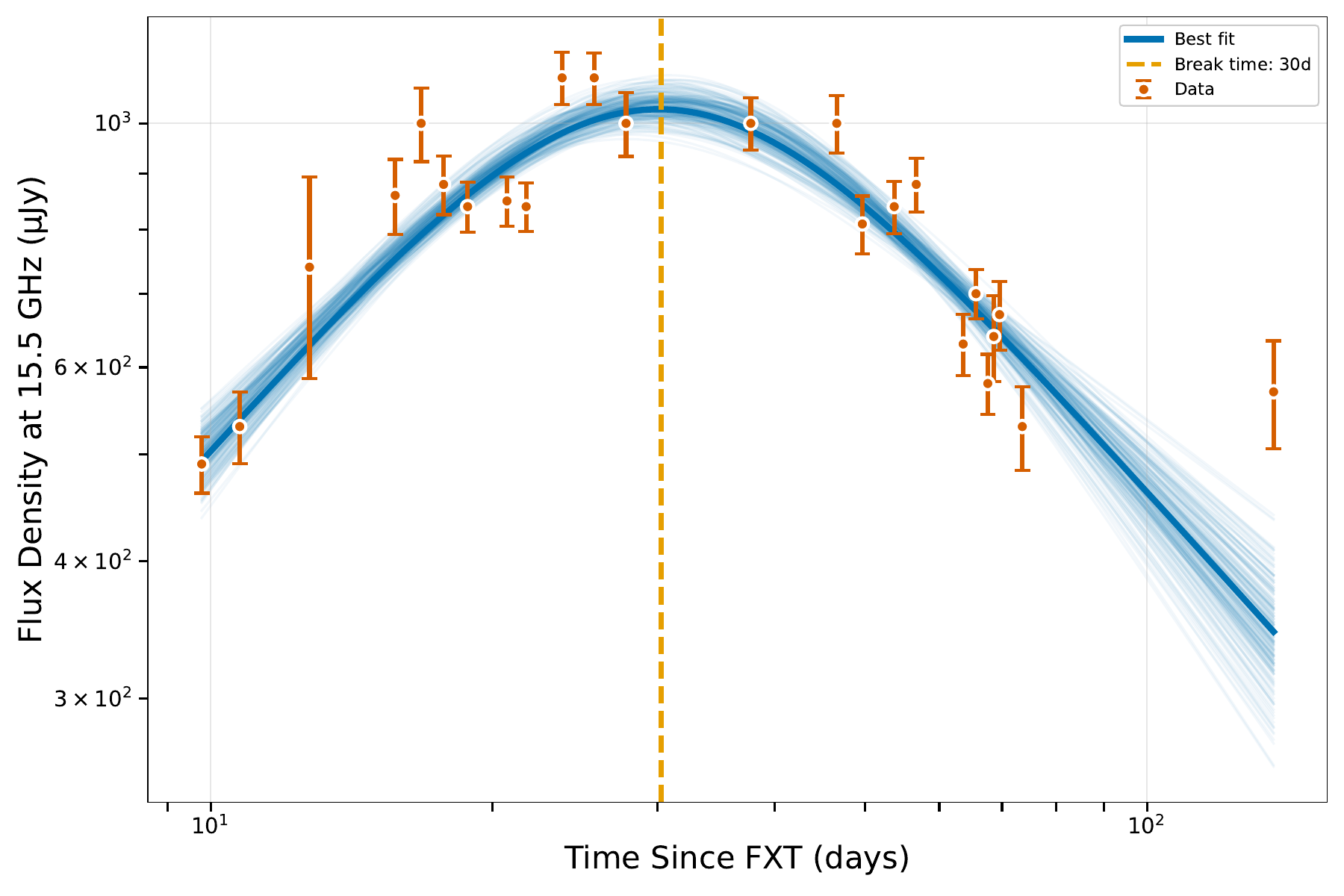}
    \caption{A time-evolving broken power law fit to all detections across the AMI-LA monitoring campaign of EP241021a. The indices for the best-fit rise and decline phases are $\alpha_{2} = 0.96 \pm 0.10 $ and $\alpha_{3} = -0.95 \pm 0.15$ respectively, peaking at $\sim$\,30 days with $F_p \sim 1030.0\,\muup\rm{Jy}$ at a central frequency of 15.5\,GHz. We include a transparent overlay of model realizations of 200 random samples from the MCMC posterior chains.}
    \label{fig:ep241021A_fitting}
\end{figure}

Our radio observations of EP241021a sample the radio afterglow at 15.5\,GHz for half a year following the explosion. We detect clearly evolving radio emission consistent with expected synchrotron behaviour for sources of this class, with the hallmark behaviour of a rise and fall with higher frequencies peaking earlier in time compared with studies of other frequencies \citep{shuEP241021a, yadav_radio_2025, gianfagnaSoftXrayTransient2025a}, and a single-peaked temporal curve. The detections are recorded in \Cref{tab:AMI-LA}, and plotted in \Cref{fig:ep241021A_fitting}. To measure temporal indices and determine the break frequency of the radio light curve, we fit the EP241021a data using a model-agnostic time-evolving broken power law of the form

\begin{equation}\label{eq:broken_single_freq}
    F(\nu,t)=F_0\bigg[0.5\bigg(\frac{t}{t_{b}}\bigg)^{-s\alpha_{2}} + 0.5\bigg(\frac{t}{t_{b}}\bigg)^{-s\alpha_{3}}\bigg]^{-1/s}
\end{equation}

where $F(\nu,t)$ is the flux density, $F$ is the amplitude of the broken power law, $t$ is the time since the FXT trigger, $t_{b}$ is the break time, and $\alpha_{i}$ are the time indices of the amplitude, rise, and fall, and s determines the smoothness of the transition between the two power-law regimes. We fit the data with \Cref{eq:broken_single_freq} using the Python package \texttt{emcee}, a flexible implementation of the MCMC technique. We fix the smoothness parameter s to be 2, with uniform priors $0 < F_0 < 2000$, $20 < t_b < 100$, $0 < \alpha_2 < 4$, and $-3 < \alpha_3 < 0$.The results of fitting \Cref{eq:broken_single_freq} to all detections of EP241021a across our AMI-LA monitoring are shown in \Cref{fig:ep241021A_fitting}. 

From the best-fitting model parameters, we can extract the break time and peak flux density of the SED at 15.5\,GHz. The light curve peaks at $t_{b} = 30.1 \pm 2.8$\,d after the FXT trigger in the observer frame, with a peak flux density of $F_\nu = 1030.0 \pm 2.8 \,\muup\rm{Jy}$. Temporal indices for the rise and decay are $\alpha_{2} = 0.96 \pm 0.10 $ and $\alpha_{3} = -0.95 \pm 0.14$ respectively. We report all uncertainties at 68\% confidence levels.

\subsection{EP240315a}\label{sec:results_EP240315a}

\begin{figure}
	\includegraphics[width=\columnwidth]{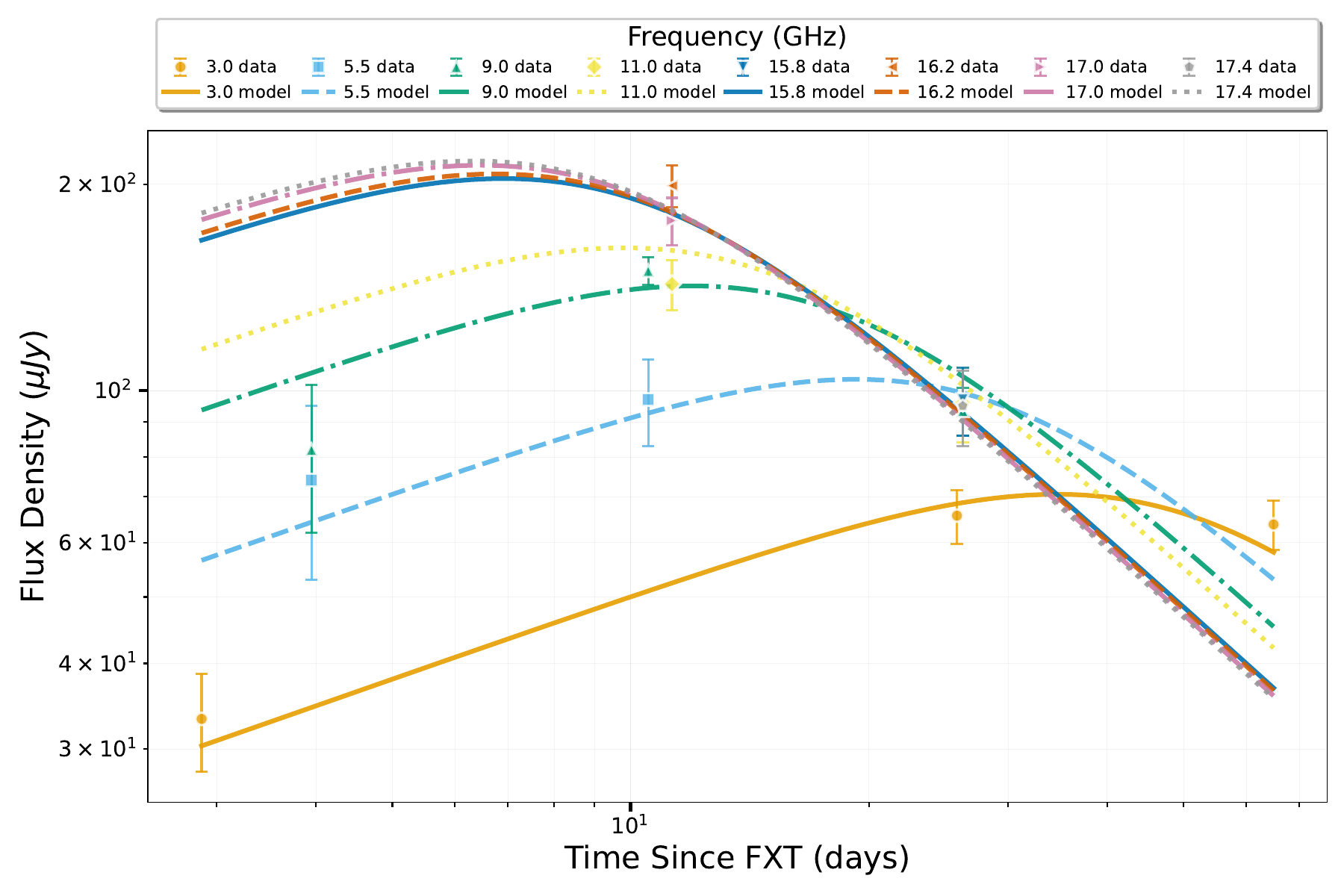}
    \caption{Flux curves of the transient EP240315a for frequencies from 3.0\,GHz to 17.4\,GHz in the sample, extracted from the fitting of \Cref{eq:broken_multi_freq} to the full dataset. The model and fitting parameters described in \Cref{sec:results_EP240315a} demonstrates the time evolution of the radio light curves with one rise index and one fall index, scaling inversely with frequency towards broader and later peaks. A subset of the data and corresponding fit curves are plotted here for clarity.}
    \label{fig:ep240315a_MCMC}
\end{figure}

The EP240315a detections added by this work provide flux densities at higher radio frequencies between 11.0 and 17.5\,GHz at 11.2 and 26.3 days post-explosion in the observer frame, improving the frequency and time sampling of the total coverage reported in the literature so far. We combine the VLA data presented in this work with the flux densities presented in other published works \citep{carotenuto2024,gillanders2024,liu2024,ricciLongtermRadioMonitoring2025}. We again fit the full set of EP240315a observations using a modified form of \Cref{eq:broken_single_freq}, which allows us to take advantage of our multi-frequency sampling by fitting a broken model with one set of temporal indices to multiple sets of light curves at distinct frequencies:

\begin{equation}\label{eq:broken_multi_freq}
    F(\nu,t)=F_{p}(t)\bigg[0.5\bigg(\frac{\nu}{\nu_{b}}\bigg)^{-s\alpha_{2}} + 0.5\bigg(\frac{\nu}{\nu_{b}}\bigg)^{-s\alpha_{3}}\bigg]^{-1/s}
\end{equation}

where $F_{p}(t) = F_0(t/t_b)^{\alpha_1}$ is the time-evolving amplitude of the power law and $\nu_{b}(t) = \nu_{0}(t/t_{b})^{\alpha_{4}}$ is the time-dependent break frequency. The results of the fitting are shown in \Cref{fig:ep240315a_MCMC}.

% We fix $t_{b}$ to be the epoch where we have greatest frequency coverage, at $\Delta T=11.279\,\rm{d}$.

We note that the $9\,\rm{GHz}$ ATCA observation at $\Delta T=19.576\,$ days since the explosion \citep{liu2024} appears abnormal when compared to our VLA observations at $\Delta T=11.256\,\rm{d}$ and $\Delta T=26.288\,\rm{d}$. A flux evolution between the two observations of $150\,\muup\rm{Jy}$ at the distance of EP240315a ($z=4.859\pm0.002$; \citealt{gillanders2024}) at $9\,\rm{GHz}$ would imply a minimum brightness temperature of $T_{B}\sim5\times10^{16}\,\rm{K}$ which is large even when considering beaming effects (e.g. \citealt{pietka2015} find, based on radio variability, that GRBs typically have $13\lesssim\rm{log}T_{B}\lesssim16$). This, coupled with the drastic change in spectral index (implying $\alpha=1.7$, i.e. suddenly optically thick at these frequencies) not seen in any of our radio observations leads us to discard this measurement and exclude it from our modelling. 

We use priors selected uniformly in the ranges $0 < F_0 < 1000$, $-3 < \alpha_1 < 3$, $1.5 < \nu_0 \textless 20$, $0.5 < \alpha_2 < 4$, $-3 < \alpha_3 < 0$, $-2 < \alpha_4 < 2$, and again fix the smoothness $s = 2$.We fix the normalization parameter $t_b$ to be the epoch for which we have the greatest frequency coverage, $t_b = 11.28$\,days post-explosion in the observer frame, which results in a break frequency at $t_b$ of $\nu_0 = 13 \pm 2$\,GHz. We find temporal indices for the rise and decay of $\alpha_{2} = 1.03 \pm 0.11 $ and $\alpha_{3} = -0.38 \pm 0.15$, respectively, with errors indicating 68\% confidence intervals. We find a break evolution index of $\alpha_4 = -1.03 \pm 0.13$, and a flux density evolution index of $\alpha_1 = -0.66 \pm 0.06$ with a peak amplitude at $t_b$ of $F_0 = 174.0 \pm 1.1 \,\muup\rm{Jy}$. 

\section{Discussion}\label{sec:discuss}

\subsection{Minimum Constraints on Energy and Lorentz Factors}\label{sec:constraints} 

With a well-sampled light curve and a clear flux density peak, it becomes possible to extract approximate minimum constraints on the internal energy release, brightness temperature, and expansion speed of an emitting region, provided that the size of the emitting region is known and that the emission can be adequately described by a single impulsive ejection involving one population of relativistic electrons expanding from a compact region, with no evidence for multiple overlapping emission components. In certain AGN systems the size of the synchrotron-emitting region is directly measurable, which allows the magnetic field and size to be constrained independently \citep{scottreadhead1977a}. For unresolved transient events such as those considered here, the synchrotron self-absorption (SSA) turnover instead provides the dominant constraint linking the flux density, magnetic field, and emitting radius. These estimates rely on the assumption that the flare is synchrotron self-absorbed at its spectral peak, such that the optical depth at the turnover frequency satisfies $\tau_{v_p} \simeq 1$.

We follow the standard equipartition approach, wherein the emitting region is modelled as a single, homogeneous component containing a population of non-thermal electrons with a power-law distribution and a uniform magnetic field (see, e.g. \citealt{fender_synchrotron_2019, matsumotoGeneralizedEquipartitionMethod2023}). Under these assumptions, the optically thick synchrotron flux at the peak frequency depends on the apparent solid angle of the emitting region and the magnetic field strength, and the total energy of a synchrotron source exhibits a pronounced minimum where magnetic field energy and electron energy reach approximate equipartition. For synchrotron self-absorption fields, magnetic and particle energies additionally scale sensitively with the source radius $R$ in opposing ways, resulting in a strong minimum as a function of source size. The minimum occurs very close to equipartition and links the source radius, magnetic field strength, and SSA peak flux density such that the synchrotron equations can be recast to depend on observable parameters--luminosity, spectral properties, and redshift, and the temporal evolution of the emission--rather than requiring prior knowledge of the physical size of the emitting region.

Following \citet{matsumotoGeneralizedEquipartitionMethod2023}, we can calculate the Newtonian equipartition energy and radius as

\begin{equation}\label{eq:min_energy}
    E_{\text{eq,N}} = 6.2 \times 10^{49} d_{L,28}^{40/17} F_{\nu,\text{mJy}}^{20/17} \nu_{10}^{-1} (1 + z)^{-37/17} \text{ erg } 
\end{equation}

\begin{equation}\label{eq:radius}
    R_{\text{eq,N}} = 1.9 \times 10^{17} d_{L,28}^{16/17} F_{\nu,\text{mJy}}^{8/17} \nu_{10}^{-1} (1 + z)^{-25/17} \text{ cm } 
\end{equation}

where $F_\nu$ and $\nu_{10}$ are the peak flux density in units of mJy and the observing frequency in units of 10\,GHz respectively, $d_{L}$ is the luminosity distance to the source in units of $10^{28}$\,cm, and $z$ is the redshift.

In the relativistic regime, normalizing the actual radius and energy as $r \equiv R/R_{\text{eq,N}}$ and $e \equiv E/E_{\text{eq,N}}$, the energy becomes

\begin{equation}\label{eq:rel_energy}
e(r, \Gamma, \theta) = \Gamma \delta_{\text{D}}^{-43/17} \left[ \frac{11}{17} \left( \frac{r}{\Gamma \delta_{\text{D}}^{-7/17}} \right)^{-6} + \frac{6}{17} \left( \frac{r}{\Gamma \delta_{\text{D}}^{-7/17}} \right)^{11} \right]
\end{equation}

and no longer has a global minimum due to the additional variables of Lorentz factor ($\Gamma$) and viewing angle ($\theta$), requiring an additional constraint to determine the radius and energy. An appropriate constraint arises from the relationship between the observed time since explosion and the emitting radius once relativistic kinematics and light-travel-time effects are taken into account. For a relativistically expanding source, the observed time satisfies
\begin{equation}
t_{\rm obs} \sim \frac{R}{\beta c\,\Gamma\,\delta_{\rm D}},
\end{equation}
implying that the emitting radius is no longer an independent parameter for a fixed observing epoch. Again following \citet{matsumotoGeneralizedEquipartitionMethod2023}, this leads to the constraint $r = \left( \frac{\beta}{\beta_{\text{eq,N}}} \right) \Gamma \delta_{\text{D}}$.

To apply the relativistic equations, it becomes necessary to quantify the expansion velocity and the relativistic effects. The apparent expansion velocity, derived under synchrotron self-absorption and equipartition conditions, can be expressed in terms of observable quantities as 

\begin{equation}\label{eq:expansion_velocity}
    \beta_{\text{eq,N}} = \frac{(1 + z)R_{\text{eq,N}}}{ct} \approx 0.73 \left[ \frac{F_{p,\text{mJy}}^{\frac{8}{17}} d_{L,28}^{\frac{16}{17}} \eta^{\frac{3}{5}}}{\nu_{p,10}(1 + z)^{\frac{8}{17}}} \left( \frac{t}{100 \text{ days}} \right)^{-1} \right] f_A^{\frac{7}{17}} f_V^{-\frac{1}{17}}
\end{equation}

\noindent where $f_A$ and $f_V$ are area and volume filling factors respectively, $\eta$ is a value dependent on the ordering of the self-absorption frequency $\nu_{sa}$ and the emitting frequency of the minimal energy electron in the synchrotron distribution $\nu_{m}$, and all other parameters depend on the position of the peak at a given time and observing frequency. If we assume that $\nu_{sa} \textgreater \nu_{m}$ and the spectral peak is due to self-absorption, $\eta$ is unity. For a fixed Lorentz factor, the energy is minimized for on-axis configurations ($\theta$ = 0), with minimal energy decreasing as $e_{min} \propto \Gamma^{-26/17}$ for increasingly relativistic sources, while off-axis viewing angles require orders of magnitude more energy \citep{matsumotoGeneralizedEquipartitionMethod2023}. If the on-axis expansion speed can be found, the minimum energy of the source at the time of observation can be determined. With the apparent expansion velocity, following \citet{brightRadioCounterpartFast2025}, the on-axis jet speed corresponding to the relativistic minimum energy condition at equipartition can be found by numerically solving the approximation

\begin{equation}\label{eq:beta_on}
\beta_{\text{eq,N}}^{\frac{17}{12}} = \beta_{\text{on}}^{\frac{17}{12}} \left[ \frac{1 + \beta_{\text{on}}}{1 - \beta_{\text{on}}} \right]
\end{equation}

for the on-axis expansion velocity $\beta_{\text{on}}$. The resulting energy and radius calculated from observables can then be considered approximate minima.

\subsection{EP241021a}\label{sec:discuss_ep241021A}

% Comparison with other papers and conclusions that can be made

EP241021a, like the majority of EP-FXTs \citep{yadav_radio_2025, oconnorRedshiftDistributionEinstein2025}, has no prompt gamma-ray detection down to deep limits \citep{svinkin2024}, despite its luminous X-ray flare and multiwavelength emission. The transient exhibits a bright, red optical counterpart \citep{busmann_curious_2025}, coincident with the X-ray and radio positions. Follow-up radio observations reveal a long-lived afterglow extending $\sim\,100$\,days post discovery, with sustained radio emission and clear temporal evolution across multiple frequency bands, which fits well within the range of counterpart luminosities expected from long GRBs and relativistic TDEs. EP241021a's comparatively low redshift of $z \approx 0.75$ makes it unlikely that a GRB counterpart would have been shifted out of the gamma-ray regime, and implies either no GRB association or an intrinsically very faint gamma-ray counterpart.

% Applying the equations of the previous section: energy and radius constraints

Applying \Cref{eq:min_energy} to the inferred peak values derived in \Cref{sec:results}, we derive a lower limit on the Newtonian explosion energy of $E_{eq,N} = 2.8 \times 10^{49}$\,ergs, and a Newtonian radius of $R_{eq,N} = 7.6 \times 10^{16}$\,cm. In the relativistic case, we can apply \Cref{eq:expansion_velocity} with the assumption that the peak is in the optically thick regime of the spectrum; that is, $\nu_m\,\textless\,v_{sa}$, where $\nu_m$ is the frequency of the minimal electron in the energy distribution of the system and $\nu_{sa}$ is the self-absorption frequency below which the spectrum is optically thick. In this regime, where the spectral peak is due to self-absorption, $\eta = 1$. As we lack information on the source structure and the source is not resolved, we further take $f_A$ and $f_V$ to each be 1. We note that while the exact values of $\eta$, $f_A$, and $f_V$ depend on the emission regime, varying them does not significantly change the result of the equation as they are constrained oppositely, as $\eta \geq 1$ and $f_A,f_V \leq 1$. With these conditions, we obtain an apparent velocity of $\beta_{\text{eq,N}}$ = 1.71.

Solving \Cref{eq:beta_on} numerically for the on-axis expansion speed, we find $\beta_{\text{on}}$ = 0.62 and therefore a minimum on-axis Lorentz factor of $\Gamma_{\text{on}} \sim 1.27$. This result suggests that the outflow is likely only mildly relativistic at $t_{b} \sim 30.2$\,days post-FXT. Taking $\beta_{\text{on}}$ to be the "true" expansion speed for the minimum energy case, we apply the constraint $r = \left( \frac{\beta}{\beta_{\text{eq,N}}} \right) \Gamma \delta_{\text{D}}$ (see \Cref{sec:constraints}) to obtain a normalized minimum expansion radius. Substituting into \Cref{eq:rel_energy}, we derive a non-normalized limit on the relativistic total energy of $E_{\text{on}} = 5.8 \times 10^{48}$\,erg, at a radius of $R_{\text{on}} = 7.2 \times 10^{16}$\,cm. 

% General gamma constraints and extra figure

\begin{figure}
	\includegraphics[width=\columnwidth]{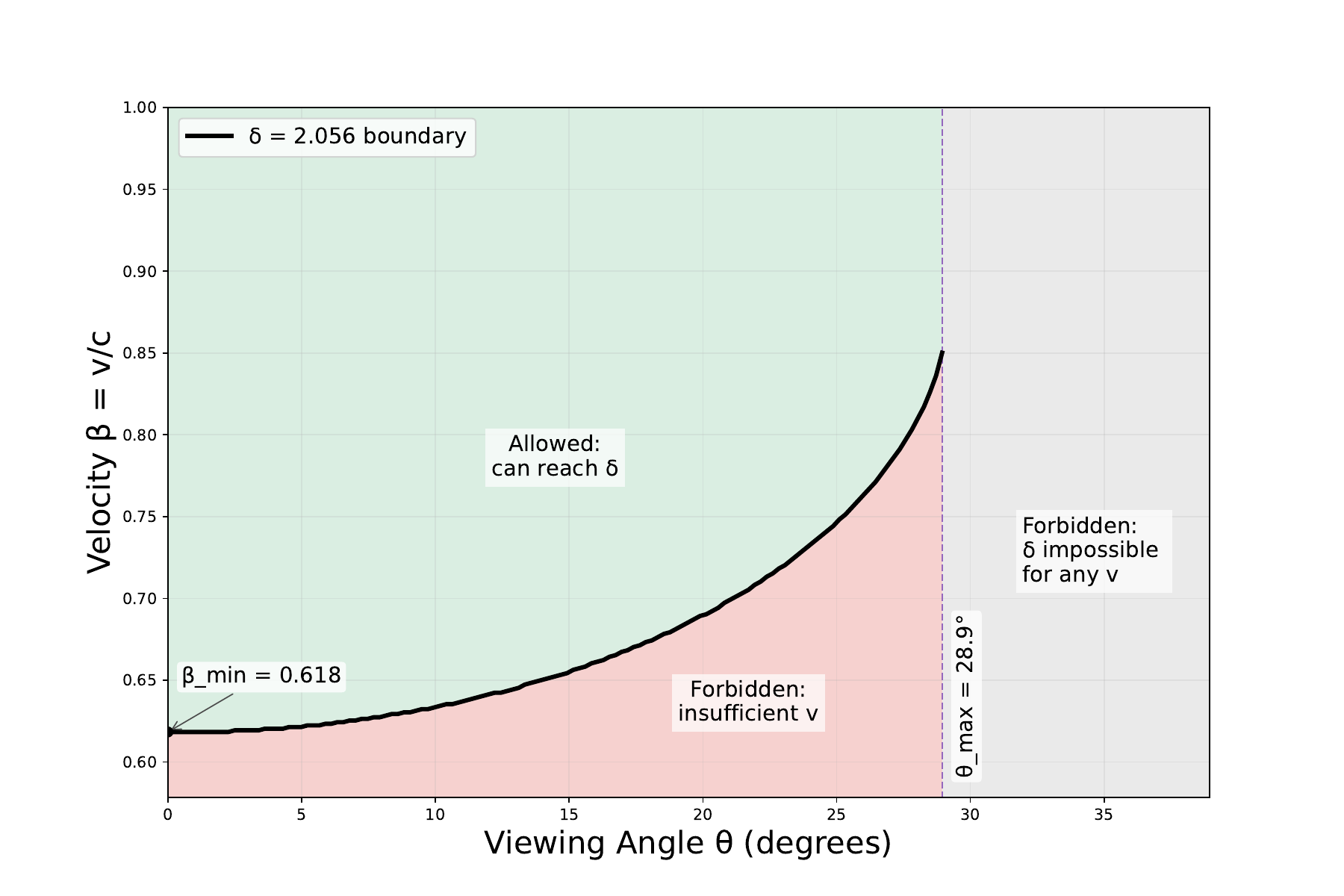}\caption{Relativistic jet parameter space for the Doppler factor $\delta_D = 2.056$ inferred for EP241021a. The solid black curve marks the boundary in $(\beta, \theta)$ space that satisfies the Doppler-factor relation $\delta_D = [\Gamma(1 - \beta \cos\theta)]^{-1}$ for a given intrinsic velocity $\beta = v/c$ and viewing angle $\theta$. Combinations of $\beta$ and $\theta$ above this curve (green region) can reproduce the required Doppler factor, while combinations below it (red region) cannot reach $\delta_D = 2.056$ for any physically allowed velocity. The minimum permitted speed is $\beta_{\min} \approx 0.618$, and solutions exist only up to a maximum viewing angle of $\theta_{\max} = 28.9^{\circ}$. This boundary therefore defines the full set of jet configurations consistent with the observed apparent expansion speed of EP241021a; in all cases, $\beta_{\rm on}$ should be regarded as a lower limit.}

    \label{fig:ep241021A_gridsearch}
\end{figure}

In the general case, any combination of jet angle and rest frame speed must satisfy the physical constraints imposed by relativistic beaming. We perform a grid search to determine the allowed combinations of $\theta$ and $\beta$ that satisfy the physical constraints imposed by relativistic beaming. Since equipartition yields a minimum expansion velocity $\beta_{\text{on}}$, the corresponding Doppler factor calculated directly from the minimum expansion velocity is $\delta_{\text{D}} = 2.056$, and represents the minimum value consistent with the energetics of the source. Any combination of viewing angle $\theta$ and off-axis $\beta$ must therefore satisfy $\delta_{\text{D}} \geq 2.1$ to remain physical. \Cref{fig:ep241021A_gridsearch} shows the resulting boundary as a function of viewing angle and $\beta$; strongly relativistic speeds are only required for the widest opening angles, with the allowed parameter space requiring a maximum viewing angle of $28.9\degree$ and a minimum speed of $0.62c$.

% Conclude here or save for later section?

The radio properties of EP241021a indicate an at least mildly relativistic outflow, consistent with a jetted origin. The equipartition analysis yields $\beta_{\mathrm{on}} = 0.62 \,(\Gamma_{\mathrm{on}} \approx 1.27)$ and an apparent expansion speed $\beta_{\mathrm{eq,N}} = 1.71$, allowing viewing angles up to $\theta_{\max} = 28.9^{\circ}$. Such parameters naturally describe a relativistic jet viewed moderately off-axis, in which the beaming of prompt high-energy emission is reduced below current detection limits \citep{Granot_2002}. The smooth rise and decline of the 15.5\,GHz light curve, peaking at $\sim$30\,days, are consistent with afterglow emission from an outflow that decelerates as its more energetic core gradually enters the line of sight \citep{rossiAfterglowLightCurves2002,Beniamini2023}.

\citet{shuEP241021a} suggest that the optical rebrightening of EP241021a may arise from the emergence of an angularly structured jet viewed off-axis, in which slower, wider-angle material dominates early emission while the jet core becomes visible at later times. \citet{gianfagnaSoftXrayTransient2025a} develop a similar geometry in detail, modeling the optical and X-ray light curves with wide, low-Lorentz-factor jet wings and an off-axis core; however, they attribute the radio afterglow to a separate, mildly relativistic quasi-spherical cocoon rather than to the jet itself, based on the high self-absorption frequency and bell-shaped radio spectra. \citet{yadav_radio_2025} explore both a structured jet and a highly relativistic off-axis jet. In the former framework the radio emission would trace the same outflow responsible for the late-time optical brightening, whereas in the latter it arises from material at larger polar angles; we note that the expansion speed of the cocoon component in \citet{gianfagnaSoftXrayTransient2025a} is comparable to our equipartition result, so the radio data alone may not discriminate between these possibilities. Our inferred minimum energy of $E_{\text{on}} = 5.8 \times 10^{48}$\,erg and mildly relativistic dynamics fall within the range expected for lower-luminosity, moderately off-axis afterglows discussed in the literature \citep[e.g.][]{ghirlanda_compact_2019, Beniamini2023}. Several studies have shown that such low-energy or mildly relativistic jets can arise from a variety of physical scenarios—including off-axis viewing and, in some cases, increased baryon loading that reduces the attainable Lorentz factor \citep[e.g.][]{beniaminiRobustFeaturesOffAxis2022}—without necessarily requiring the highly relativistic, multi-component structure invoked in these works, which \citet{shuEP241021a} found difficult to reconcile with the optical behaviour. Within the current uncertainties, the radio observations are therefore consistent with a moderately off-axis, mildly relativistic jet, aligning EP241021a with the broader population of low-luminosity or off-axis GRB afterglows, while more detailed modelling will be required to distinguish among the possible physical origins of the outflow.

\subsection{EP240315a}

\begin{table}
\centering
\caption{Summary of relativistic velocity and bulk Lorentz factor constraints derived from our multi-frequency peak analysis of EP240315a. The values of $\beta_{\rm on}$ and $\beta_{\rm eq,N}$ are computed using \Cref{eq:expansion_velocity} and \Cref{eq:rel_energy} respectively (see \Cref{sec:constraints}. The monotonic decrease in all quantities with increasing peak time is consistent with deceleration of the outflow, noting that these 
remain lower limits. All quantities are reported in the observer frame and represent lower limits for an on-axis jet, evaluated at each spectral peak extracted from our fits at the corresponding observing frequency (see \Cref{sec:results_EP240315a}.)}
\label{tab:beta_analysis}
\begin{tabular}{ccccc}
\hline
Frequency & Peak Time & $\beta_{eq,N}$ & $\beta_{\textrm{on}}$ & $\Gamma$ \\
(GHz) & (days) & & & (Lorentz) \\
\hline
17.39 & 6.31 & 16.8  & 0.965 & 3.84 \\
17.00 & 6.45 & 16.7 & 0.965 & 3.83 \\
16.24 & 6.75 & 16.5 & 0.965 & 3.80 \\
15.85 & 6.91 & 16.4 & 0.964 & 3.78 \\
14.70 & 7.43 & 16.0 & 0.963 & 3.73 \\
14.32 & 7.62 & 15.9 & 0.963 & 3.72 \\
13.93 & 7.83 & 15.8 & 0.963 & 3.70 \\
13.16 & 8.27 & 15.6 & 0.962 & 3.66 \\
12.78 & 8.51 & 15.4 & 0.962 & 3.64 \\
12.40 & 8.76 & 15.3 & 0.961 & 3.62 \\
11.00 & 9.84 & 14.8 & 0.959 & 3.55 \\
9.00 & 11.95 & 14.1 & 0.957 & 3.43 \\
5.50 & 19.24 & 12.3 & 0.948 & 3.16 \\
5.00 & 21.10 & 12.0 & 0.947 & 3.11 \\
3.00 & 34.60 & 10.5  & 0.937 & 2.86 \\
\hline
\end{tabular}
\end{table}

% Applying the equations of the previous section: energy and radius constraints
The observations of EP240315a sample multiple frequencies over a significant segment of the radio afterglow's duration. Some frequencies, such as those at 3.0\,GHz, 5.5\,GHz, and 9.0\,GHz are sampled on both the rising and falling slopes (shown in \Cref{fig:ep240315a_MCMC}). Using the parameters from the fit of \Cref{eq:broken_multi_freq}, we calculate peaks, apparent expansion speeds, on-axis expansion speeds, and on-axis Lorentz factors for all observing frequencies. The results are shown in \Cref{tab:beta_analysis}. The deceleration from $\sim$\,5 days (in the observer frame) to $\sim$\,65 days post-FXT can be clearly seen, as well as the significantly relativistic minimum speeds of the jet outflow. The peak emission characteristics and expansion speeds at each frequency suggest a minimum energy limit of $E_{\text{on}} = 8.1 \times 10^{48}$\,erg at 6.5 days evolving to a minimum energy limit of $E_{\text{on}} = 3.6 \times 10^{48}$\,erg at our last measurable peak at 34.6 days, taking the same assumptions described in \Cref{sec:constraints}. Under the assumption that $\beta_{\textrm{on}}$ is the true velocity, we find a deceleration rate of $t^{-0.18}$, slightly slower than the theoretical deceleration rate of $t^{-3/8}$ for a jet in a homogeneous medium \citep{piranGammarayBurstsFireball1999a}. 

\begin{figure*}
\vspace*{-0.5cm}
	\includegraphics[width=\textwidth]{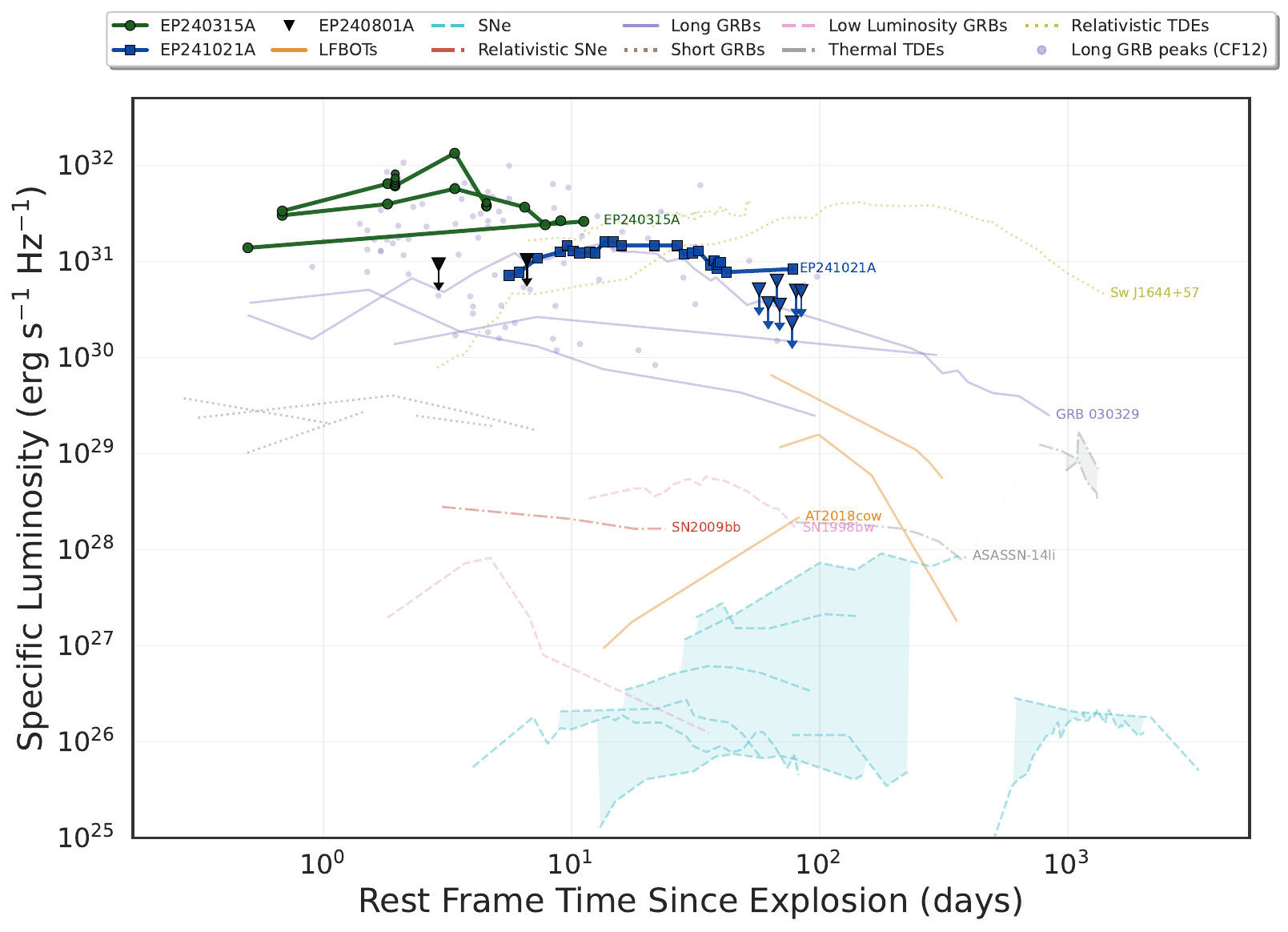}
    \caption{Rest-frame radio light curves of Einstein Probe fast X-ray transients compared with representative classes of extragalactic transients. Solid markers and lines show the radio luminosity evolution of the three EP FXTs with detected counterparts or luminosity limits and known redshifts in our sample: EP240315a (blue circles), EP241021a (green circles for detections, green triangles for $3\sigma$ upper limits), and EP240801a (black triangles for $3\sigma$ upper limits). EP240618a, which lacks a redshift, appears only in the flux-space comparison (\Cref{fig:ata_flux_limits}). Luminosities are expressed as rest-frame specific luminosity ($L_{\nu}$) versus rest-frame time since explosion. Background curves show individual well-sampled transients, with archetypes labeled; shaded bands span the supernova sample of \citet{bietenholzRadioLuminosityrisetimeFunction2021} and the tidal disruption event sample of \citet{cendesUbiquitousLateRadio2024}, and small circles mark the peak luminosities of the long-GRB radio afterglows of \citet{chandra_radio-selected_2012}. The EP-FXT radio counterparts occupy a luminosity range overlapping the brightest GRB and relativistic TDE afterglows, consistent with moderately to highly relativistic outflows.}
    \label{fig:luminosity_comparison}
\end{figure*}

\begin{figure}
	\includegraphics[width=\columnwidth]{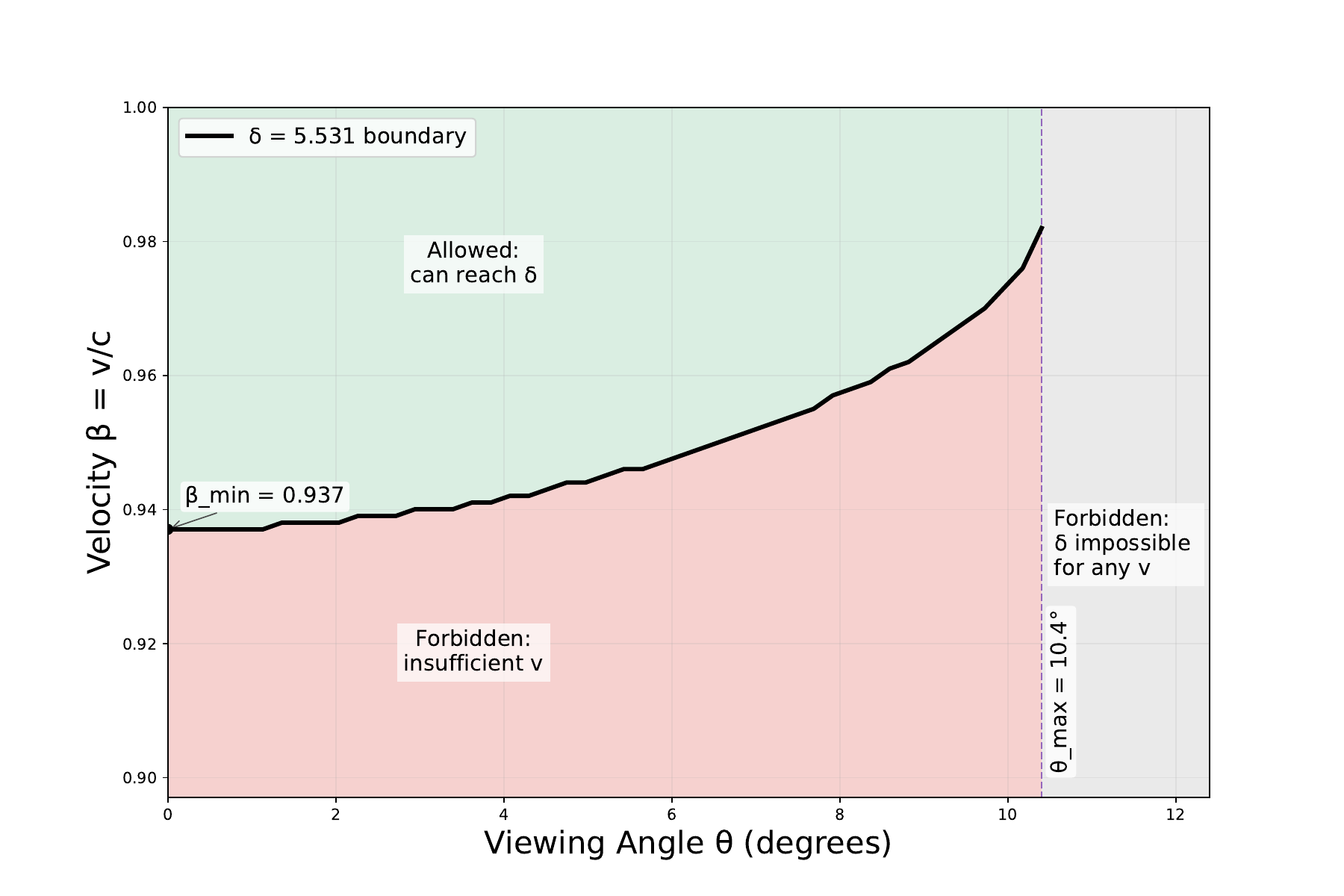}\caption{Relativistic jet parameter space for the Doppler factor $\delta_D = 5.531$ inferred for EP240315a at 3.0\,GHz. The solid black curve marks the boundary in $(\beta, \theta)$ space that satisfies the Doppler-factor relation $\delta_D = [\Gamma(1 - \beta \cos\theta)]^{-1}$ for a given intrinsic velocity $\beta = v/c$ and viewing angle $\theta$. Combinations of $\beta$ and $\theta$ above this curve (green region) can reproduce the required Doppler factor, while combinations below it (red region) cannot reach $\delta_D = 5.5315$ for any physically allowed velocity. The minimum permitted speed is $\beta_{\min} \approx 0.937$, and solutions exist only up to a maximum viewing angle of $\theta_{\max} = 10.4^{\circ}$. This boundary defines the full set of jet configurations consistent with the observed apparent expansion speed of EP240315a; again, in all cases, $\beta_{\rm on}$ should be regarded as a lower limit.}

    \label{fig:ep240315a_gridsearch}
\end{figure}

The constraints reported in \Cref{tab:beta_analysis}, as discussed, represent the on-axis case. Given the already-high on-axis expansion speeds required to produce the observed Doppler shift, it is unlikely for the EP240315a emission to be at a wide viewing angle to us. To quantify this, we again consider the general case, choosing the best-constrained peak at 3.0\,GHz. We find that only on-axis solutions with viewing angles less than 10.4$\degree$ and highly relativistic speeds greater than 0.937$c$ are required with our calculated equivalent expansion speed (\Cref {fig:ep240315a_gridsearch}). 

With the spectral information available across the observations of EP240315a, it becomes possible to extend the analysis in \Cref{sec:discuss_ep241021A} to multiple frequencies. Radio afterglows from GRBs are typically interpreted in the context of a fireball afterglow model, where the broadband SED is described by a series of power-law segments (PLSs) separated by break frequencies. We can analyse the evolution of the radio spectral energy distribution in the context of GRB models which predict how characteristic break frequencies and their flux density evolve with time. For radio frequencies the key breaks are the synchrotron self-absorption frequency ($\nu_{sa}$), the minimum electron energy frequency ($\nu_m$), and the cooling frequency ($\nu_c$) which together define a twice- or thrice- smoothly broken power-law spectrum as outlined in \citet{granotShapeSpectralBreaks2002}. We extract spectral indices for the optically thick and optically thin slopes from our fitted model. The spectral index evolves from $\alpha = 1.03 \pm 0.11$ to $\alpha = - 0.38 \pm 0.15$. Higher frequencies, while fitted well by the power law in \Cref{eq:broken_multi_freq}, are only sampled at one data point. Evaluating the spectrum in context of the afterglow spectra presented by \citet{granotShapeSpectralBreaks2002}, the spectral index at early times is flatter than most PLSs but is closest to segment C, which occurs in fast cooling regimes when $\nu_{sa} \textgreater \nu_{ac}$ and $\nu_{m} \textgreater \nu_c$. Our spectral index suggests that at these times, our sampled frequencies may span a transition region, wherein $\nu_{sa}$ is evolving towards lower frequencies, or that $\nu_{sa}$ lies somewhere within our observing bands. The spectrum gradually flattens, then inverts at $\Delta T=\sim\,35\,\rm{d}$. The spectrum remains too flat to allow for segments H (implied $p = 0.72$) and G (implied $p = 1.72$). We do not measure a fully optically thick slope (i.e., $\alpha \sim 2$–2.5), suggesting we are unlikely to be observing a single, wholly self-absorbed segment of the spectrum. Instead, the relatively shallow spectral index may indicate that our observing bands lie within a transition region between optically thick and optically thin emission, or that we observe multiple populations of electrons. It is also possible that the self-absorption and cooling breaks are located relatively close together in frequency, such that multiple falling spectral segments are not individually resolved. In that case, as we fit only one break, we would effectively observe a single apparent break from optically thin to optically thick emission, smoothed by our fitting. Furthermore, the observed temporal evolution of the break frequency ($\alpha_4 = -1.03 \pm 0.13$) is inconsistent with the \citet{granotShapeSpectralBreaks2002} Spectrum 1 prediction of $\nu_{sa} \propto t^{-3/5}$, providing additional evidence that our observations do not correspond to a single well-defined spectral segment. Nevertheless, given the frequencies we observe and that we observe evolution from thin to thick over the observing period, our observations are most consistent with spectrum 1 or spectrum 2.

We note that the late-time evolution may be poorly constrained due to only one late-time observation, which may result in the underprediction of uncertainties due to degeneracy in the original light curve fitting, indicated by strong correlation between $\nu_b$ and $\alpha_4$. Spectral indices should therefore be interpreted with appropriate caution. 

Taken together, the multi-frequency light-curve evolution, inferred expansion speeds, and broadband spectral behaviour of EP240315a indicate that the radio afterglow arises from a relativistic, jetted outflow. The observed deceleration from highly relativistic apparent velocities, the on-axis Lorentz factors derived from our modelling, and the evolution of the spectral index within the fast-cooling regime are consistent with expectations for synchrotron emission from a long-duration GRB afterglow. Although the late-time spectrum is less well constrained due to limited sampling, the available data are well described within the framework of standard GRB afterglow physics, supporting the interpretation of EP240315a as a canonical long GRB event.

\subsection{Survey Limits}\label{sec:limits}

%\textcolor{red}{Either sampling the GRB distribution and assigning redshifts to our upper limits, or drawing from the distribution and seeing if we are likely to detect any sources?}

% Sampling of redshifts to generate luminosity constraints for GRB sources? See if the fluxes of the general GRB population is consistent with 

\begin{figure*}
    \includegraphics[width=\textwidth]{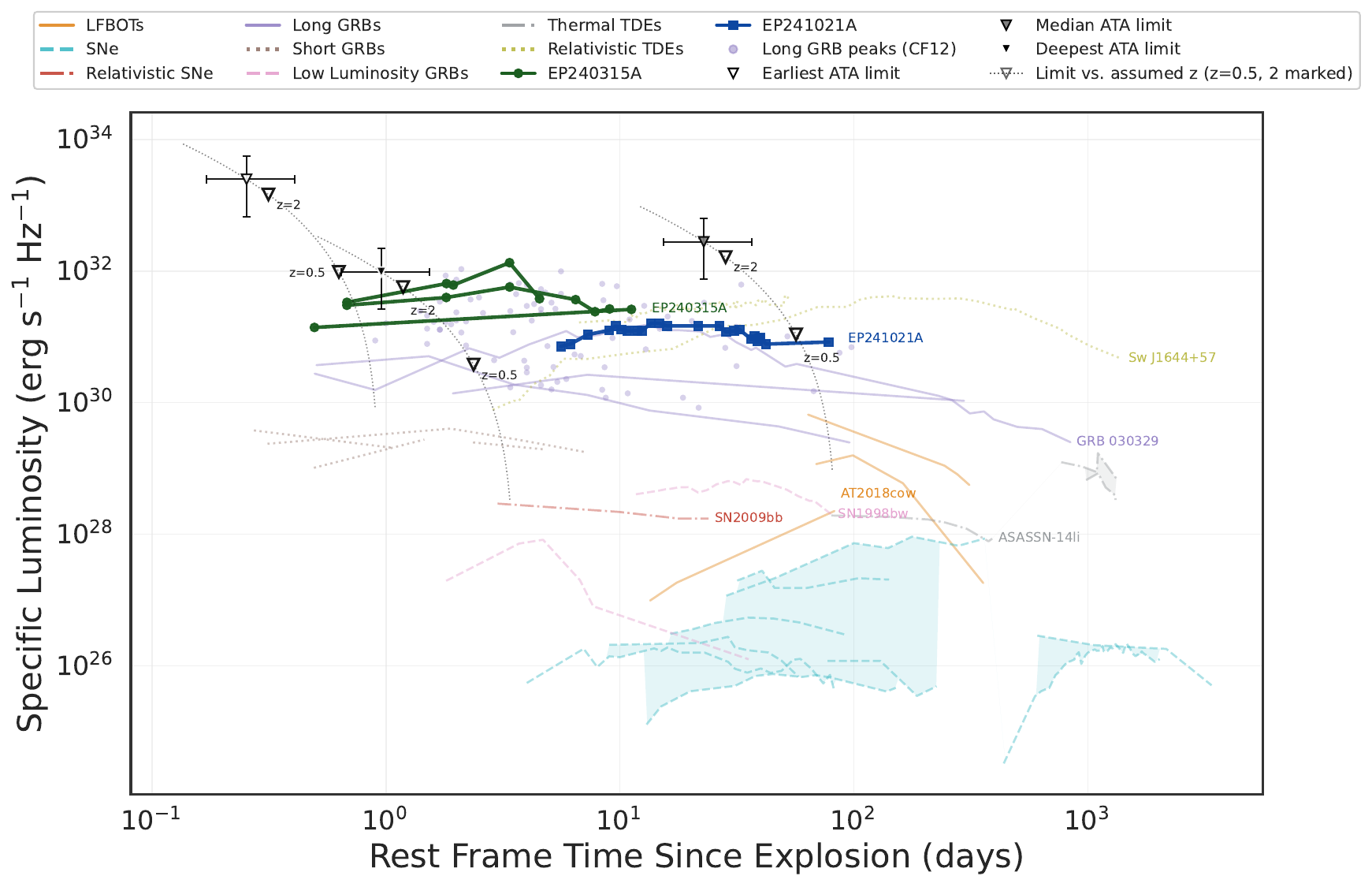}
    \caption{ATA limits mapped into the rest-frame $(t_{\rm rest}, L_{\nu})$ plane under a GRB-like redshift prior. We sample redshifts from the observed GRB distribution of \citet{pescalli2016} and show only the earliest, median-sensitivity, and deepest ATA limits for clarity. Crosshairs mark the median over the redshift prior with 16--84\% ranges in both axes. Dotted curves trace each limit through the plane as a function of assumed redshift, with open triangles marking the limit at fixed $z=0.5$ and $z=2$. The radio light curves of the two EP FXTs with detected counterparts, EP240315A ($z=4.8$) and EP241021A ($z=0.748$), are shown in the foreground. Representative transient light curves are plotted in luminosity units for context, with shaded bands spanning the supernova sample of \citet{bietenholzRadioLuminosityrisetimeFunction2021} and the tidal disruption event sample of \citet{cendesUbiquitousLateRadio2024}, and small circles marking the peak luminosities of the long-GRB afterglows of \citet{chandra_radio-selected_2012}.}
    \label{fig:grb_redshift_lims}
\end{figure*}
Without redshifts or knowledge of a detailed afterglow model, the upper limits obtained with the ATA and AMI-LA are most usefully interpreted in the context of known objects. To quantify the sensitivity of the limits and survey completeness for one relevant class of objects, we calculate detection fractions by Monte-Carlo sampling from the GRB redshift distribution presented in \citet{pescalli2016}, which assumes the GRB rate tracks the cosmic star-formation history. This distribution provides a realistic empirical prior on the distances at which cosmological GRBs occur. For each ATA epoch $i$ with flux-density limit $F_{\nu,i}^{\rm lim}$, we draw $N = 1000$ independent redshift samples $z_j$ from this prior. For each sampled redshift we compute the corresponding specific-luminosity threshold implied by the flux-density limit,
\begin{equation}
    L_{\nu,{\rm lim},i}(z) \;=\; \frac{4\pi D_L^2}{(1+z)}\,F_{\nu,i}^{\rm lim},
\end{equation}
where the factor of $(1+z)^{-1}$ follows the standard flux–luminosity relation for specific luminosity evaluated at the observer-frame frequency. We explicitly omit any K-correction, following the guidance of \citet{hoggDistanceMeasuresCosmology2000} for non-bolometric corrections: this omission is equivalent to assuming a flat spectral index, but avoids imposing an arbitrary or potentially misleading spectral assumption across a heterogeneous set of sources with unknown redshifts. Accordingly, the resulting luminosities should be interpreted as observer-frame specific luminosities, each corresponding to a different rest-frame frequency for each sampled redshift.

This procedure yields, for each epoch, a distribution of 1000 possible luminosity thresholds reflecting the range of distances at which a GRB could plausibly occur. For any trial intrinsic luminosity $L_\nu^\star$, we then estimate the fraction of GRB-like redshifts at which a source of that luminosity would have been detected. This ``detection fraction'' is computed as
\begin{equation}
    f_i(L_\nu^\star) \;=\; \Pr_{z}\!\big[L_\nu^\star > L_{\nu,{\rm lim},i}(z)\big],
\end{equation}
representing the fraction of sampled redshifts that produce a luminosity greater than $L_{\nu,{\rm lim},i}(z)$ for a given $F_{\nu,i}^{\rm lim}$. In other words, for each flux-density limit, there exists an observer-frame luminosity threshold $L_{\nu,90,i}$ above which a transient would have been detectable for 90\% of redshifts drawn from the GRB-rate prior. We refer to these values as the 90\% observer-frame specific-luminosity limits associated with each flux-density constraint. This statistic expresses our constraints directly in terms of intrinsic luminosity as observed at the ATA frequency, marginalized over the physical redshift distribution expected for cosmological GRBs. We note that these limits implicitly assume that the transient would have been observed near its peak radio luminosity; GRB afterglows that were still rising or already fading at the time of observation could be intrinsically brighter than our quoted limits yet remain undetected.

\begin{figure}
    \includegraphics[width=\columnwidth]{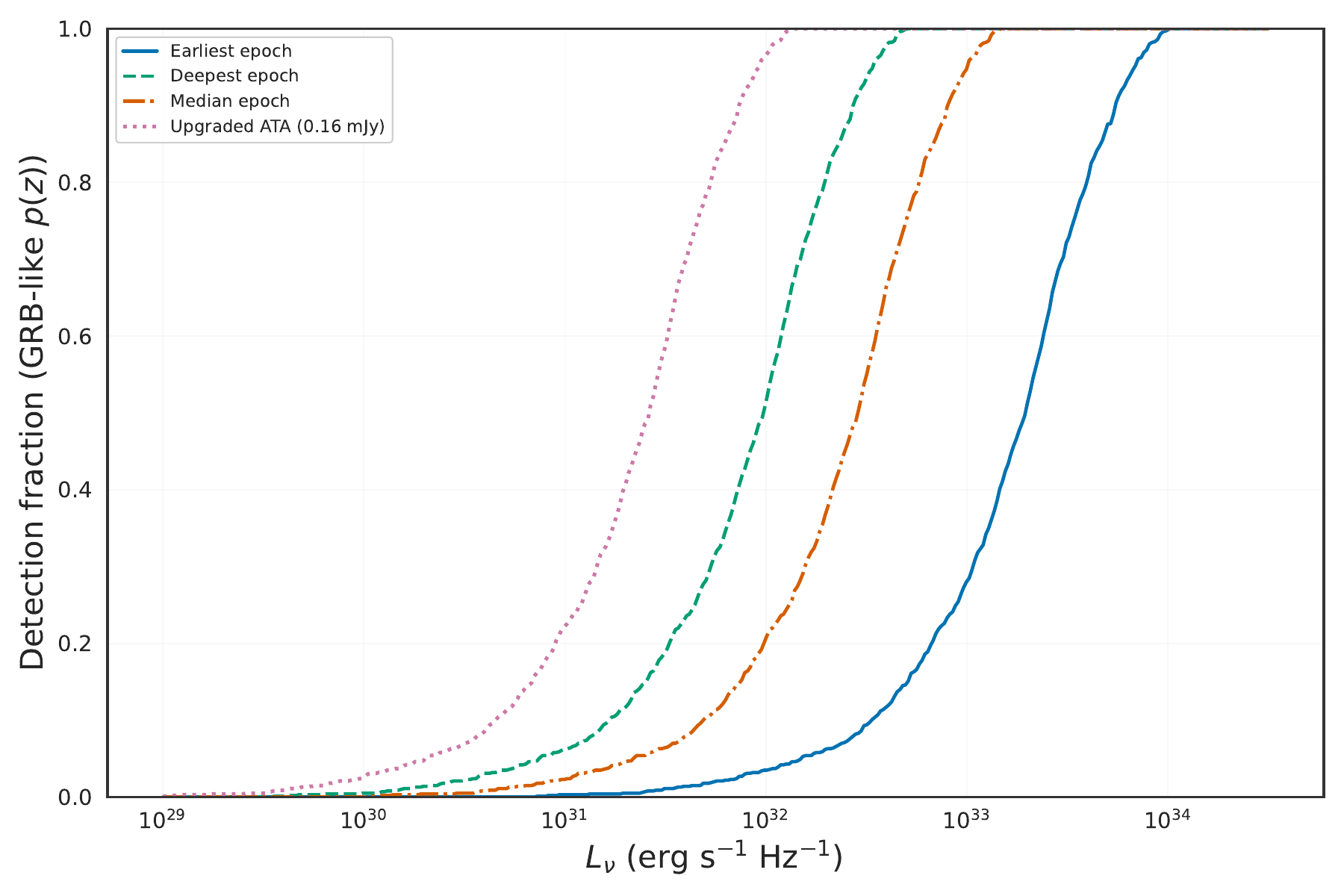}
    \caption{Detection fraction versus intrinsic specific luminosity $L_{\nu}$ under the GRB redshift prior of \citet{pescalli2016}. Curves are shown for the earliest, median sensitivity, and deepest ATA limits (as in \Cref{fig:grb_redshift_lims}). We include the limit that will be possible with the upgraded ATA (see \Cref{sec:facilities}).}
    \label{fig:detection_fraction}
\end{figure}

\Cref{fig:grb_redshift_lims} presents the earliest (post-explosion), median-sensitivity, and deepest limits mapped into this observer-frame luminosity space. Because we omit K-corrections, we report only specific luminosities at the ATA observing frequency. \Cref{fig:detection_fraction} displays detection-fraction curves for the same limits, as well as the projected limits attainable with the upgraded ATA for comparable integration times. These curves provide a model-independent characterization of the program's completeness as a function of intrinsic $L_{\nu}$. The earliest epoch represents a constraint on 90\% of possible luminosities of EP240626a at 3340\,MHz at 0.94 observer-frame days post-explosion, and the deepest epoch represents a constraint on EP240913a at 3340\,MHz at 3.55 observer-frame days post-explosion.

Assuming the same GRB-like redshift distribution, our $3\sigma$ ATA non-detections imply observer-frame specific-luminosity limits of 
$L_{\nu} < 2.5 \times 10^{33}$, $2.8 \times 10^{32}$, and $9.7 \times 10^{31}\ \mathrm{erg\ s^{-1}\ Hz^{-1}}$ 
for the \emph{earliest}, \emph{median-sensitivity}, and \emph{deepest} observations, respectively (medians of the 16--84\% ranges; see \Cref{fig:grb_redshift_lims}). Interpreted in this observer-frame context, these limits lie above the bulk of long- and short-GRB radio afterglows and are comparable to the bright end of relativistic TDEs; thus typical cosmological GRBs remain allowed and only the most luminous sources would be detectable \citep{piranGammarayBurstsFireball1999a}. We find no evidence for an extreme radio-rebrightening episode exceeding the brightest known counterparts at comparable phases under this distance prior. This conclusion is consistent with the detection statistics of \citet{chandra_radio-selected_2012}, who showed that only $\sim30$\% of GRBs are detected in the radio even with $\sim100~\mu$Jy sensitivities at 8.5\,GHz (their Table~2), and that most detections cluster within a relatively narrow luminosity range (their Fig.~4). The ALARRMS project \citep{andersonArcminuteMicrokelvinImager2018} showed the detection rate of radio counterparts can reach $\sim 50\%$ with ($ \textless\,1$\,day) follow-up, indicating a higher possible rate of detection for those early measurements, though the majority of our limits cover a range of times post-burst more similar to the sample of \citet{chandra_radio-selected_2012}. Although the ATA observations are at different frequencies than either paper's sample, their flux-density distributions and detection fractions provide useful context; our limits naturally fall above the bulk of this population, reinforcing the expectation that only unusually luminous GRB afterglows would be detectable in our sample. For nearby scenarios ($z \lesssim 0.2$), the same flux limits would correspond to $L_{\nu} \lesssim 10^{30-31}\ \mathrm{erg\ s^{-1}\ Hz^{-1}}$, which begins to constrain long-GRB afterglows and relativistic TDE counterparts (also see \Cref{fig:luminosity_comparison}).

% the third image complements the other two (three representative crosshair limits in luminosity space and the detection probability curves) by disaggregating the result to the level of individual observations within the program. We may not want to include all of these plots but we can remove or include as we want to

\begin{figure*}
    \centering
    \includegraphics[width=\textwidth]{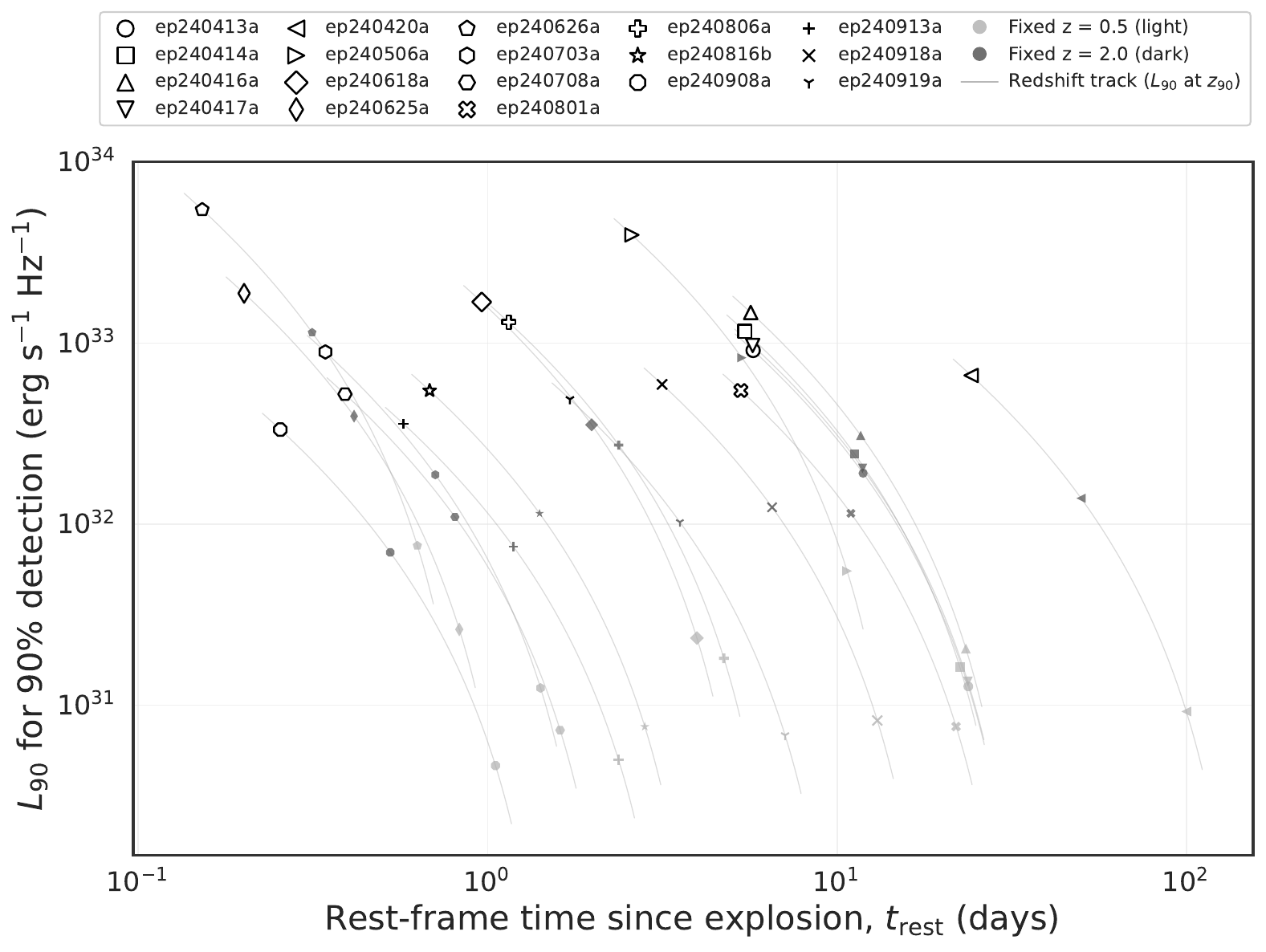}
    \caption{Rest-frame 90\% detection-probability limits derived from the
    earliest ATA limit at 2660\,MHz for each source. Open symbols each
    correspond to one source, showing the characteristic rest-frame
    time since explosion ($t_{\mathrm{rest}}$) and the luminosity
    threshold $L_{90}$ above which 90\% of Monte Carlo--sampled
    GRB redshifts would yield a detectable signal given the measured
    flux-density limit. The limits are computed using the GRB-like
    redshift prior adopted throughout this work, such that $L_{90}$
    represents the luminosity at which our program would have detected
    a GRB-like transient at 90\% of probable redshifts. 
    For comparison, each source's limit is also evaluated at two fixed
    redshifts, $z=0.5$ and $z=2.0$, plotted with the same symbol as the
    source in light and dark gray respectively. All three markers for a
    given source lie on a common redshift track (thin line), the locus
    $L_{\nu}(z)=4\pi D_L^2(z)\,F_{\nu}/(1+z)$ versus
    $t_{\rm rest}=t_{\rm obs}/(1+z)$ traced as the assumed redshift
    varies. The $L_{90}$ marker is placed at $z_{90}$, the
    90th-percentile redshift of the GRB prior, so that both of its
    coordinates derive from a single redshift and it lies on the same
    physical track as the fixed-$z$ points; because $L_{\nu}$ increases
    monotonically with $z$, this $L_{90}$ equals the 90th percentile of
    the luminosity distribution over the prior. At fixed redshift
    the limiting luminosity is not a probabilistic threshold but a concrete luminosity limit.}

    \label{fig:ata_L90_trest}
\end{figure*}

To summarize the luminosity constraints achieved by the ATA monitoring campaign, we compute for every source and epoch the intrinsic specific luminosity $L_{90}$ that would yield a 90\% detection probability under the same GRB-like redshift prior used throughout this section. Each epoch is then placed in the $(t_{\rm rest},L_\nu)$ plane at a representative rest-frame time (the median of $t_{\rm obs}/(1+z)$ over the sampled redshifts) and labelled by the associated source. The resulting distribution (shown in \Cref{fig:ata_L90_trest}) provides a per-source, per-epoch view of the campaign’s coverage across the $\sim$1–65~day window sampled: points at lower $L_{90}$ identify the epochs with the deepest limits, while the spread with time reflects variations in cadence and completeness of the afterglow sampling.

% Discussion of luminosities if things were at a distance/redshift of blank, statistics if you can find a valid method

\subsection{Future Facilities}\label{sec:facilities}

\begin{figure*}
\vspace{-0.5cm}
	\includegraphics[width=\textwidth]{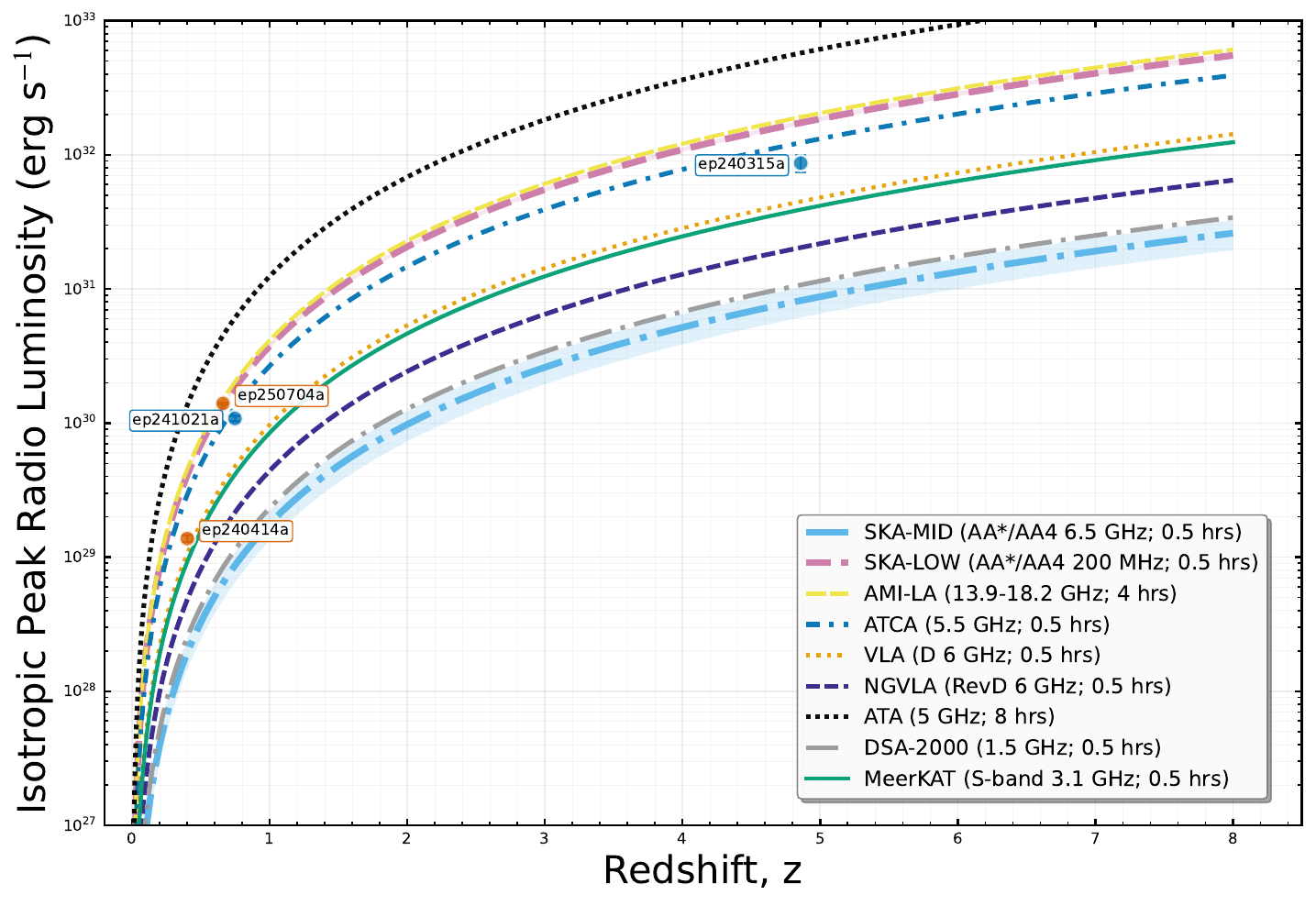}
    \caption{Comparison of the ATA sensitivity with current and next-generation radio facilities. Curves show the limiting isotropic peak radio luminosity as a function of redshift for representative 1000\,s integrations at the indicated observing frequencies. The ATA sensitivity curve (black dotted) reflects the performance during the early half of this survey (21 antennas, dual-tuning configuration). The DSA, SKA-LOW (green shaded; AA* and AA4 configurations), SKA-MID (gold shaded; AA* and AA4 configurations) \citep{SKAOSensitivityCalculator}, AMI-LA \citep{zwart2008}, MeerKAT \citep{SARAOApps}, ng-VLA \citep{NgvlaEct}, ATCA \citep{ATCAcalc}, DSA \citep{DSA}, and VLA \citep{VLAExposureCalculator} are shown for comparison. The DSA curve closely follows the SKA-MID sensitivity but at lower frequencies (1–2 GHz). Points mark radio discovery luminosities on Einstein Probe FXTs with known redshifts and radio counterparts.}
    \label{fig:facilities}
\end{figure*}

To date, the EP-FXTs for which radio counterparts have been discovered are still rare, and radio counterparts have been discovered at $\sim$0.1\,mJy levels by sensitive facilities with fairly short ($\sim$ few hours long) observations. In \Cref{fig:facilities} we present luminosity curves of the sensitivities of current and future facilities against the luminosities of heretofore discovered radio counterparts, scaling each to observation lengths typical or expected for observations scheduled with each facility.

Planned developments to the ATA will substantially improve both bandwidth and sensitivity over the coming years. During this survey, the correlator backend was upgraded to digitize four simultaneous tunings, providing up to $\sim$2,688\,MHz of processed bandwidth with flexible placement across 1–11\,GHz. Further expansion of backend compute capacity is expected to permit even broader contiguous frequency coverage. In parallel, the development and deployment of new quadridge feeds, designed for improved broadband impedance matching and substantially lower system temperatures, will reduce thermal noise across the full operating range, enabling more sensitive wideband observations. The array is also being expanded from 28 to its full complement of 43 active antennas, increasing the effective collecting area and thus theoretical sensitivity by 50\%. Combined with the doubling of processed bandwidth, these enhancements will yield sensitivity improvements of $\sim1.5\sqrt{2}$ relative to the initial configuration used during this survey. The resulting instrument will offer significantly deeper imaging capability and wider simultaneous bandwidth for transient monitoring and rapid-response observations.

Relative to the limits achieved in this ATA campaign, the next generation of widefield centimeter- and meter-wave interferometers will extend continuum sensitivity by more than an order of magnitude, reaching flux densities below a few $\mu$Jy within comparable integration times. As illustrated in \Cref{fig:luminosity_comparison}, the Square Kilometre Array (SKA) will provide sub-$\mu$Jy sensitivities across both its mid- and low-frequency components, enabling detections of GRB- and TDE-like afterglows to redshifts $z\gtrsim5$. The Deep Synoptic Array–2000 (DSA–2000), though operating at lower frequencies ($\sim$1–2\,GHz), will achieve similar continuum sensitivity within a much wider instantaneous field of view, offering a complementary discovery space for high-cadence, all-sky transient monitoring. Its sensitivity curve coincides with that of SKA-MID in the figure, despite the distinct spectral range.

At higher frequencies, the planned Next Generation Very Large Array (ngVLA) will provide sub-arcsecond imaging and wide instantaneous bandwidth across 3–30\,GHz, bridging the capabilities of SKA-MID and existing high-frequency facilities. Together, these instruments will enable rapid, wideband observations of faint extragalactic transients that would be inaccessible to current facilities. Sensitivities that presently require several hours of integration with the ATA will be achieved in seconds, permitting systematic, population-scale studies of relativistic outflows and radio afterglows over cosmological volumes. These advances will place future targeted surveys, such as the present one, within a broader statistical context for transient populations discovered by the next generation of widefield radio arrays. Finally, while the sensitivities of forthcoming large-collecting-area arrays will open new parameter space for faint, high-redshift transients, smaller facilities such as the ATA will continue to occupy an important niche. Their flexible scheduling, rapid-response capability, and suitability for brighter nearby events ensure that they remain valuable for prompt follow-up and targeted monitoring that does not require the full sensitivity of next-generation instruments.

\section{Conclusions}

We have presented a coordinated radio follow-up program of 20 Einstein Probe fast X-ray transients using the Allen Telescope Array (ATA), supplemented by targeted observations with MeerKAT, the Karl G. Jansky Very Large Array (VLA), ATCA, eMERLIN, and AMI-LA. From the campaign we report radio detections of EP240315a and EP241021a and flux-density measurements or $3\sigma$ upper limits for the remaining sample. The program comprises 59 radio epochs that sample the first $\sim$1--65~days after the X-ray triggers and map the mJy--sub-mJy sensitivity frontier for rapid FXT follow-up on week-to-month timescales. Under a GRB-like redshift prior, the ATA $3\sigma$ non-detections correspond to observer-frame specific-luminosity limits of $L_{\nu}\lesssim2.5\times10^{33}$, $2.8\times10^{32}$, and $9.7\times10^{31}\ \mathrm{erg\ s^{-1}\ Hz^{-1}}$ at our earliest, median, and deepest epochs (medians of the 16--84\% ranges). At $z\lesssim0.2$, those same flux limits imply $L_{\nu}\lesssim10^{30\text{--}31}\ \mathrm{erg\ s^{-1}\ Hz^{-1}}$, a regime that begins to constrain long-GRB afterglows and the brightest relativistic TDEs.

The two best-sampled radio counterparts yield physically informative constraints under standard synchrotron and equipartition assumptions. For EP241021a, modelling of the AMI-LA 15.5~GHz light curve with a time-evolving broken power law gives a peak at $t_b\approx30$~days; with equipartition analysis under relativistic minimum-energy conditions, we find a minimum on-axis speed $\beta_{\rm on}\simeq0.62$ ($\Gamma_{\rm on}\approx1.3$) and a minimum on-axis energy of order $E_{\rm on}\sim5.8\times10^{48}$\,erg. For EP240315a, multi-frequency modelling indicates highly relativistic early behaviour followed by measurable deceleration; equipartition estimates give minimum on-axis energies of order $10^{48\text{--}49}$~erg on timescales of several days and a deceleration law shallower than the canonical $t^{-3/8}$ expectation for a uniform medium (we infer $\propto t^{-0.18}$ from our fits). These numerical estimates assume equipartition, unity filling factors, and that the observed spectral peaks are driven by synchrotron self-absorption; relaxing those assumptions shifts the absolute values but not the qualitative inference that jetted outflows can reproduce the observed radio evolution.

Looking forward, both facility and strategy improvements will substantially increase sensitivity to FXT radio counterparts. Upgrades to the ATA (expanded processed bandwidth, improved low-noise feeds, and completion to the full 42-antenna array) are expected to yield an aggregate sensitivity gain of order $\sim1.5\sqrt{2}$ relative to the configuration used in this survey, enabling deeper and faster follow-up with the same instrument. More transformative gains will come from next-generation arrays (SKA, DSA, ngVLA), which will reach sub-$\mu$Jy continuum sensitivities on minute--hour timescales and will convert studies like the present one from targeted case studies into systematic, population-level probes of FXTs and their progenitors.

Finally, we note observational priorities motivated by our findings: (1) rapid, multi-band radio follow-up within days of X-ray detection to capture rising light curves and constrain break frequencies; (2) coordinated optical and infrared spectroscopy to secure redshifts and host types; (3) VLBI, polarization, and high-frequency radio observations to constrain source sizes, geometry, and magnetic-field structure; and (4) larger, uniform radio surveys of EP alerts to clarify the true occurrence rate of radio-bright FXTs. In the near term, the present dataset establishes a baseline for the radio properties of the newly revealed Einstein Probe FXT population and highlights both the promise and the observational challenges of connecting soft X-ray transient discoveries to their radio counterparts.

\section*{Acknowledgements}

% MeerKAT - General
The MeerKAT telescope is operated by the South African Radio Astronomy Observatory, which is a facility of the National Research Foundation, an agency of the Department of Science and Innovation.
% MeerKAT - S-band
This work has made use of the "MPIfR S-band receiver system" designed, constructed and maintained by funding of the MPI f\"{u}r Radioastronomy and the Max-Planck-Society.
% VLA
The National Radio Astronomy Observatory and Green Bank Observatory are facilities of the U.S. National Science Foundation operated under cooperative agreement by Associated Universities, Inc.
% ATA
The Allen Telescope Array (ATA) refurbishment program and its ongoing operations receive substantial support from Franklin Antonio. Additional contributions from Frank Levinson, Jill Tarter, Jack Welch, the Breakthrough Listen Initiative and other private donors have been instrumental in the renewal of the ATA. Breakthrough Listen is managed by the Breakthrough Initiatives, sponsored by the Breakthrough Prize Foundation. The Paul G. Allen Family Foundation provided major support for the design and construction of the ATA, alongside contributions from Nathan Myhrvold, Xilinx Corporation, Sun Microsystems, and other private donors. The ATA has also been supported by contributions from the US Naval Observatory and the US National Science Foundation.
% AMI-LA
We thank the staff at the Mullard Radio Astronomy Observatory for carrying out observations with the AMI-LA. 

\section*{Data Availability}
All data used in this survey are available in tables within the paper, and images can be made available upon request.

\bibliographystyle{mnras}
\bibliography{article_bib}

\appendix

\section{ATA FXT Follow-Up Limits}\label{app:limits}

\onecolumn
\begin{longtable}[c]{cccccc}
\textbf{Date}      & \textbf{Source}    & \textbf{Frequency (MHz)} & $\Delta \textbf{T}$ & \textbf{Flux Limits (mJy)}    \\
\hline
\endfirsthead
\endhead
05012024  & ep240420a & 5000            & 10.82      & \textless 1.89       \\
05012024  & ep240420a & 8500            & 10.82      & \textless 1.02       \\
05012024  & ep240420a & 1500            & 10.66      & \textless 3.89       \\
05012024  & ep240420a & 3000            & 10.66      & \textless 1.74       \\
05022024  & ep240414a & 1500            & 17.73      & \textless 11.1       \\
05022024  & ep240414a & 8000            & 17.79      & \textless 1.04       \\
05022024  & ep240414a & 3000            & 17.73      & \textless 3.02       \\
05022024  & ep240414a & 5000            & 17.79      & \textless 2.94       \\
05042024  & ep240413a & 5000            & 20.77      & \textless 3.11       \\
05042024  & ep240413a & 8500            & 20.77      & \textless 12.1       \\
05042024  & ep240413a & 1500            & 20.66      & \textless 6.26       \\
05042024  & ep240413a & 3000            & 20.66      & \textless 1.38       \\
05072024  & ep240416a & 8500            & 21.14      & \textless 9.75       \\
05072024  & ep240416a & 5000            & 21.14      & \textless 4.15       \\
05092024  & ep240506a & 8500            & 2.101      & \textless 1.58       \\
05092024  & ep240506a & 1500            & 1.986      & \textless 6.50        \\
05092024  & ep240506a & 3000            & 1.986      & \textless 1.84       \\
05092024  & ep240506a & 5000            & 2.101      & \textless 4.06       \\
05112024  & ep240417a & 5675            & 23.56      & \textless 3.48       \\
05112024  & ep240417a & 5000            & 23.56      & \textless 1.97       \\
05132024  & ep240315a & 3000            & 58.19      & \textless 2.99       \\
05132024  & ep240315a & 5000            & 58.33      & \textless 2.72       \\
05132024  & ep240315a & 1500            & 58.19      & \textless 12.2       \\
05132024  & ep240315a & 8500            & 58.33      & \textless 2.48       \\
05152024  & ep240420a & 8500            & 24.86      & \textless 1.33       \\
05152024  & ep240420a & 5000            & 24.86      & \textless 3.44       \\
05152024  & ep240420a & 3000            & 24.62      & \textless 1.89       \\
05152024  & ep240420a & 1500            & 24.62      & \textless 4.66       \\
05182024  & ep240414a & 2660            & 33.68      & \textless 2.53       \\
05182024  & ep240414a & 7340            & 33.84      & \textless 1.50        \\
05182024  & ep240414a & 3340            & 33.68      & \textless 2.48       \\
05182024  & ep240414a & 6660            & 33.84      & \textless 2.66       \\
05192024  & ep240413a & 7340            & 35.71      & \textless 2.63       \\
05192024  & ep240413a & 6660            & 35.71      & \textless 4.56       \\
05192024  & ep240413a & 3340            & 35.6       & \textless 2.24       \\
05192024  & ep240413a & 2660            & 35.6       & \textless 1.98       \\
05212024  & ep240416a & 7340            & 35.15      & \textless 7.18       \\
05212024  & ep240416a & 3340            & 35.03      & \textless 2.26       \\
05212024  & ep240416a & 2660            & 35.03      & \textless 3.21       \\
05212024  & ep240416a & 6660            & 35.15      & \textless 4.86       \\
05222024  & ep240506a & 6660            & 16.13      & \textless 3.19       \\
05222024  & ep240506a & 3340            & 16.02      & \textless 3.57       \\
05222024  & ep240506a & 2660            & 16.02      & \textless 8.61       \\
05222024  & ep240506a & 7340            & 16.13      & \textless 2.53       \\
05232024  & ep240417a & 2660            & 35.49      & \textless 2.11       \\
05232024  & ep240417a & 3340            & 35.49      & \textless 1.55       \\
06242024  & ep240618a & 3340            & 5.954      & \textless 2.37       \\
06242024  & ep240618a & 2660            & 5.954      & \textless 3.67       \\
06262024  & ep240625a & 2660            & 1.244      & \textless 4.10       \\
06262024  & ep240625a & 3340            & 1.244      & \textless 1.94       \\
06272024  & ep240626a & 6660            & 1.131      & \textless 8.79       \\
06272024  & ep240626a & 2660            & 0.9439     & \textless 11.9       \\
06272024  & ep240626a & 7340            & 1.131      & \textless 8.57       \\
06272024  & ep240626a & 3340            & 0.9439     & \textless 15.1       \\
07032024  & ep240626a & 3340            & 0.9439     & \textless 14.4       \\
07032024  & ep240626a & 2660            & 0.9439     & \textless 11.1       \\
07052024  & ep240703a & 7340            & 2.256      & \textless 1.99       \\
07052024  & ep240703a & 3340            & 2.125      & \textless 1.44       \\
07052024  & ep240703a & 2660            & 2.125      & \textless 1.94       \\
07052024  & ep240703a & 6660            & 2.256      & \textless 1.49       \\
07112024  & ep240708a & 2660            & 2.418      & \textless 1.14       \\
07112024  & ep240708a & 3340            & 2.418      & \textless 1.08       \\
07232024  & ep240625a & 7340            & 28.19      & \textless 1.68       \\
07232024  & ep240625a & 6660            & 28.19      & \textless 1.47       \\
07292024  & ep240626a & 2660            & 32.84      & \textless 4.19       \\
07292024  & ep240626a & 3340            & 32.84      & \textless 6.52       \\
07302024  & ep240506a & 6660            & 84.85      & \textless 1.87       \\
07302024  & ep240506a & 7340            & 84.85      & \textless 1.69       \\
08012024  & ep240506a & 2660            & 87.78      & \textless 1.59       \\
08012024  & ep240506a & 3340            & 87.78      & \textless 1.35       \\
08052024  & ep240703a & 7340            & 33.05      & \textless 1.39       \\
08052024  & ep240703a & 6660            & 33.05      & \textless 1.15       \\
08062024  & ep240801a & 7340            & 4.826      & \textless 1.05       \\
08062024  & ep240801a & 6660            & 4.826      & \textless 0.752      \\
08082024  & ep240801a & 7000            & 6.806      & \textless 1.33       \\
08082024  & ep240801a & 3200            & 6.806      & \textless 2.44       \\
08122024  & ep240703a & 2660            & 40.07      & \textless 1.35       \\
08122024  & ep240703a & 3340            & 40.07      & \textless 1.07       \\
08132024  & ep240806a & 3340            & 7.113      & \textless 3.4        \\
08132024  & ep240806a & 2660            & 7.113      & \textless 2.84       \\
08162024  & ep240806a & 6660            & 10.12      & \textless 1.22       \\
08162024  & ep240806a & 7340            & 10.12      & \textless 1.72       \\
08202024  & ep240816b & 3340            & 4.224      & \textless 0.718      \\
08202024  & ep240816b & 7340            & 4.441      & \textless 1.18       \\
08202024  & ep240816b & 2660            & 4.224      & \textless 1.19       \\
08202024  & ep240816b & 6660            & 4.441      & \textless 1.0        \\
08262024  & ep240625a & 2660            & 62.11      & \textless 1.29       \\
08262024  & ep240625a & 3340            & 62.11      & \textless 1.05       \\
08272024  & ep240625a & 6660            & 63.1       & \textless 0.993      \\
08272024  & ep240625a & 7340            & 63.1       & \textless 1.26       \\
09022024  & ep240801a & 7340            & 31.89      & \textless 0.831      \\
09022024  & ep240801a & 6660            & 31.89      & \textless 0.762      \\
09022024  & ep240806a & 6660            & 27.27      & \textless 1.06       \\
09022024  & ep240806a & 7340            & 27.27      & \textless 1.27       \\
09032024  & ep240801a & 3340            & 32.85      & \textless 0.98       \\
09032024  & ep240801a & 2660            & 32.85      & \textless 1.19       \\
09032024  & ep240806a & 2660            & 28.24      & \textless 0.715      \\
09032024  & ep240806a & 3340            & 28.24      & \textless 0.662      \\
09102024  & ep240908a & 2660            & 1.58       & \textless 0.725      \\
09102024  & ep240908a & 3340            & 1.58       & \textless 0.671      \\
09132024  & ep240908a & 6660            & 4.552      & \textless 0.658      \\
09132024  & ep240908a & 7340            & 4.552      & \textless 0.64       \\
09172024  & ep240913a & 2660            & 149.5      & \textless 0.781      \\
09172024  & ep240913a & 3340            & 149.5      & \textless 0.591      \\
09172024  & ep240420a & 3340            & 3.707      & \textless 0.995      \\
09172024  & ep240420a & 2660            & 3.707      & \textless 1.44       \\
09302024  & ep240919a & 3340            & 10.65      & \textless 0.847      \\
09302024  & ep240919a & 2660            & 10.65      & \textless 1.07       \\
10012024  & ep240919a & 6660            & 11.65      & \textless 1.21       \\
10012024  & ep240919a & 7340            & 11.65      & \textless 1.87       \\
10072024  & ep240918a & 6660            & 18.78      & \textless 0.898      \\
10072024  & ep240918a & 7340            & 18.78      & \textless 1.21       \\
10082024  & ep240918a & 2660            & 19.55      & \textless 1.29       \\
10082024  & ep240918a & 3340            & 19.55      & \textless 1.41       \\
11032024  & ep241021a & 7340            & 12.99      & \textless 1.36       \\
11032024  & ep241021a & 6660            & 12.99      & \textless 1.38       \\
09102024b & ep240908a & 3340            & 1.788      & \textless 0.936      \\
09102024b & ep240908a & 2660            & 1.788      & \textless 0.907      \\
\hline
\end{longtable}

%%%%%%%%%%%%%%%%%%%%%%%%%%%%%%%%%%%%%%%%%%%%%%%%%%

% Don't change these lines
\bsp	% typesetting comment
\label{lastpage}
\end{document}